\documentclass{article}
\usepackage{graphicx}
\usepackage{epstopdf, epsfig}
\usepackage{float}
\usepackage{color}
\usepackage{soul}
\usepackage[normalem]{ulem}
\usepackage{amsmath}
\usepackage{cleveref}
\usepackage{authblk}
\usepackage{natbib}
\usepackage{subeqnarray}
\usepackage[margin=1.25in]{geometry}

\title{Density Effects on the Post-shock Turbulence Structure and Dynamics}

\author[1,*]{Yifeng Tian}
\author[1]{Farhad A. Jaberi}
\author[2]{Daniel Livescu}

\affil[1]{Department of Mechanical Engineering, Michigan State University,
	East Lansing, MI 48823, USA}
\affil[2]{CCS-2, Los Alamos National Laboratory, NM 87545, USA}
\affil[*]{Currently at CCS-2, Los Alamos National Laboratory, NM 87545, USA}

\graphicspath{{./figure/}}
\newcommand\Sc{\mbox{\textit{Sc}}} %Schmidt number
\newcommand\Pran{\mbox{\textit{Pr}}} %Schmidt number
\newcommand\Rey{\mbox{\textit{Re}}} %Schmidt number

\begin{document}
		\maketitle
		\begin{abstract}
Turbulence structure resulting from multi-fluid or multi-species, variable-density isotropic turbulence interaction with a Mach 2 shock is studied using turbulence-resolving shock-capturing simulations and Eulerian (grid) and Lagrangian (particle) methods. The complex roles density play in the modification of turbulence by the shock wave are identified.
Statistical analyses of the velocity gradient tensor (VGT) show that the density variations significantly change the turbulence structure and flow topology. Specifically, a stronger symmetrization of the joint probability density function (PDF) of second and third invariants of the anisotropic velocity gradient tensor, PDF$(Q^\ast, R^\ast)$, as well as the PDF of the vortex stretching contribution to the enstrophy equation, are observed in the multi-species case. Furthermore, subsequent to the interaction with the shock, turbulent statistics also acquire a differential distribution in regions having different densities. This results in a nearly symmetrical PDF$(Q^\ast, R^\ast)$ in heavy fluid regions, while the light fluid regions retain the characteristic tear-drop shape. To understand this behavior and the return to "standard" turbulence structure as the flow evolves away from the shock, Lagrangian dynamics of the VGT and its invariants are studied by considering particle residence times and conditional particle variables in different flow regions. The pressure Hessian contributions to the VGT invariants transport equations are shown to be not only affected by the shock wave, but also by the density in the multi-fluid case, making them critically important to the flow dynamics and turbulence structure.

	\end{abstract}
	
	\section{Introduction}
	\label{sec:intro}
	
	The interaction of a normal shock wave with multi-fluid or multi-species isotropic turbulence is an extension of the canonical Shock-Turbulence Interaction (STI) problem which includes strong variable density effects. This extended configuration can enhance our understanding of more complex flow problems such as fuel-air mixing in supersonic combustion, the interaction of supernova remnants with interstellar clouds, shock propagation through foams and bubbly liquids, Inertial Confinement Fusion (ICF), and re-shock problem in Richtmyer-Meshkov Instability (RMI). Most of the previous theoretical, numerical, and experimental studies of STI have been dedicated to the original canonical problem. 
	
	The early theoretical study by \citet{ribner1954} has restricted the STI to the linear interaction regime with a large scale separation between the shock and turbulence, so that the nonlinear and viscous effects are assumed to be negligible during the interaction. By decomposing the pre-shock turbulence into independent modes (acoustic, vortical and entropy) using Kovasznay decomposition \citep{kovasznay1953}, the post-shock turbulence statistics can be theoretically derived from the linearized Rankine-Hugoniot jump conditions. This approach is referred to as the Linear Interaction Approximation (LIA) and represents an important limiting case, since it provides analytical predictions for the jumps of fluctuating quantities across the shock. 
	
	Due to the challenges of accurate experimental measurements of the smallest time and length scales around the shock wave, numerical simulations have been widely employed to investigate this interaction. Researchers have been used both shock-capturing and shock-resolving simulations to understand the post-shock amplification of Reynolds stress, vorticity variance, and turbulent length scales \citep{lee1993,hannappel1995,mahesh1995,lee1997,mahesh1997,jamme2002,larsson2009,larsson2013}. Earlier numerical studies have shown limited agreement with the LIA predictions because the parameter range was outside the linear regime. More recently, \citet{ryu2014} have considered a wide range of parameters in their shock-resolving Direct Numerical Simulations (DNS) to show that the DNS results converge to the LIA solutions when the ratio of the shock thickness ($\delta$) to the pre-shock Kolmogorov length scale ($\eta$) becomes small. Replacing the actual shock interaction with the LIA relations can extend the reach of DNS to arbitrarily high shock Mach numbers and much larger Taylor Reynolds number ($Re_{\lambda}$) than otherwise computationally feasible, provided that the interaction parameters correspond to the linear regime. This method (named Shock-LIA by the authors) was used for detailed studies of the post-shock turbulent energy flux and vorticity dynamics \citep{livescu2015vorticity,quadros2016turbulent}. \citet{sethuraman2018thermodynamic} used shock-capturing simulation and LIA to study the thermodynamic field generated by STI. In a recent study \citep{tian2017}, we showed, using shock-capturing turbulence-resolving simulations, that the LIA predictions for the Reynolds stresses can be approached provided that the scale separation between numerical shock thickness ($\delta_n$) and Kolmogorov length scale is sufficient. Thus, when the ratio of turbulent to shock scales is large enough, so that the numerical artifacts near the shock do not influence the flow, the shock-capturing method can correctly simulate the STI.  
	
	As mentioned above, in many practical applications, STI may occur in a mixture of very different density fluids. This motivated our extension of the canonical STI problem to include variable density effects \citep{tian2017numerical,tian2017} by considering the pre-shock turbulence as an isotropic mixture of two fluids (species) with different molecular weights, as encountered in non-premixed combustion. Using turbulence-resolving shock-capturing simulations, we have examined the turbulence statistics, turbulence budgets, conditional statistics, and energy spectrum in the multi-fluid STI and found that the nonlinear effects from the density variations significantly change the turbulence properties in both physical and spectral spaces. The relation between velocity and a passive scalar field has also been studied by \citet{boukharfane2018evolution} and \citet{buttay2016analysis}. Other studies \citep{jin2015simulations, huete2017interaction} used LIA and shock-capturing simulations to study the interaction of a reactive premixed mixture with shock and turbulence. These studies help in better understanding of complex STI problem. However, there still exist many gaps in our knowledge of the variable density effects on the post-shock turbulence structure and flow topology.
	
	In this study, we focus on the density effects on the post-shock turbulence structure by examining the velocity field. The properties of the velocity gradient tensor (VGT) determine a wide variety of turbulence characteristics, such as the flow topology, deformation of material volume, energy cascade, and intermittency. Understanding both the VGT field immediately after the shock-wave and its dynamics as the flow evolves away from the shock wave is also crucial to the development of subgrid-scale models that can accurately describe the shock interaction and return-to-isotropy effects. \citet{perry1987description,chong1990general} has proposed an approach to classify the local flow topology and structure using the invariants of VGT. The dynamical behavior of the VGT has been studied for incompressible flows using the Lagrangian evolution of the invariants along conditional mean trajectories (CMT) \citep{meneveau2011lagrangian}. The statistics regarding the invariants of VGT and their Lagrangian dynamics have been used to understand the structure of turbulence in many canonical flows, such as isotropic turbulence, turbulent boundary layer and mixing layers \citep[e.g.][]{chong1998turbulence,ooi1999study,wang2012flow,bechlars2017evolution}. Previous studies on single-fluid STI have examined the PDF of VGT. \citet{ryu2014, livescu2015vorticity} took a step further to investigate the turbulence structure and vorticity dynamics based on the examination of VGT invariants. By taking the advantage of the Shock-LIA method, they extracted the statistics of VGT and its invariants for a wide range of shock Mach numbers, even though the dynamics of VGT as the turbulence evolves away from the shock wave could not be examined with the Shock-LIA method.  Our earlier numerical studies of variable density STI have revealed some important new features of velocity and scalar statistics in this setup \citep{tian2017tsfp,tian2018aiaa}.  However, these studies have not yet fully identified the variable density effects on the post-shock turbulence/scalar structure. 
	
	This study uses the recently generated database of the turbulence-resolving shock-capturing simulations of multi- and single- fluid STI to: 1) develop a better understanding of variable density and shock effects on the turbulence structure immediately after the shock wave, and 2) perform the first Lagrangian analysis of this flow configuration for better understanding of the dynamical behavior of VGT as the turbulence evolves away from the shock. While the compressibility effects are weak for the current parameter range and not discussed, variable density effects are very significant and the focus of this study. The paper is organized as follows. Details of the simulations and the testing conducted to assess the accuracy of the Lagrangian and Eulerian analysis are discussed in \cref{sec:numerical}. Results are presented in \cref{sec:results} and concluding remarks are made in \cref{sec:conclusion}.

	\section{Numerical Method and Accuracy}
	\label{sec:numerical}
	In this section, we first briefly discuss the numerical approach used for shock-capturing turbulence-resolving simulations in our previous study \citep{tian2017}, from which we have extracted the VGT statistics addressed in this paper. The extended variable-density STI configuration is described next, followed by a discussion of the new Lagrangian simulations used to examine the VGT dynamics away from the shock.
	
	\subsection{Governing Equations and Numerical approach}
	
	The conservative form of the dimensionless compressible Navier-Stokes equations for flows with two miscible species (i.e. continuity, momentum, energy, and species mass fraction transport equations) have been solved numerically together with the perfect gas law using a high-order hybrid numerical method \citep{tian2017}. The inviscid fluxes for the transport equations have been computed using the fifth-order Monotonicity Preserving (MP) scheme, as described in \citet{li2012}. The molecular transport terms have been calculated using the sixth-order compact scheme \citep{lele1992}. The 3rd-order Runge-Kutta scheme has been used for time advancement.
	
	\begin{figure*}
		\centering
		\includegraphics[width=5in]{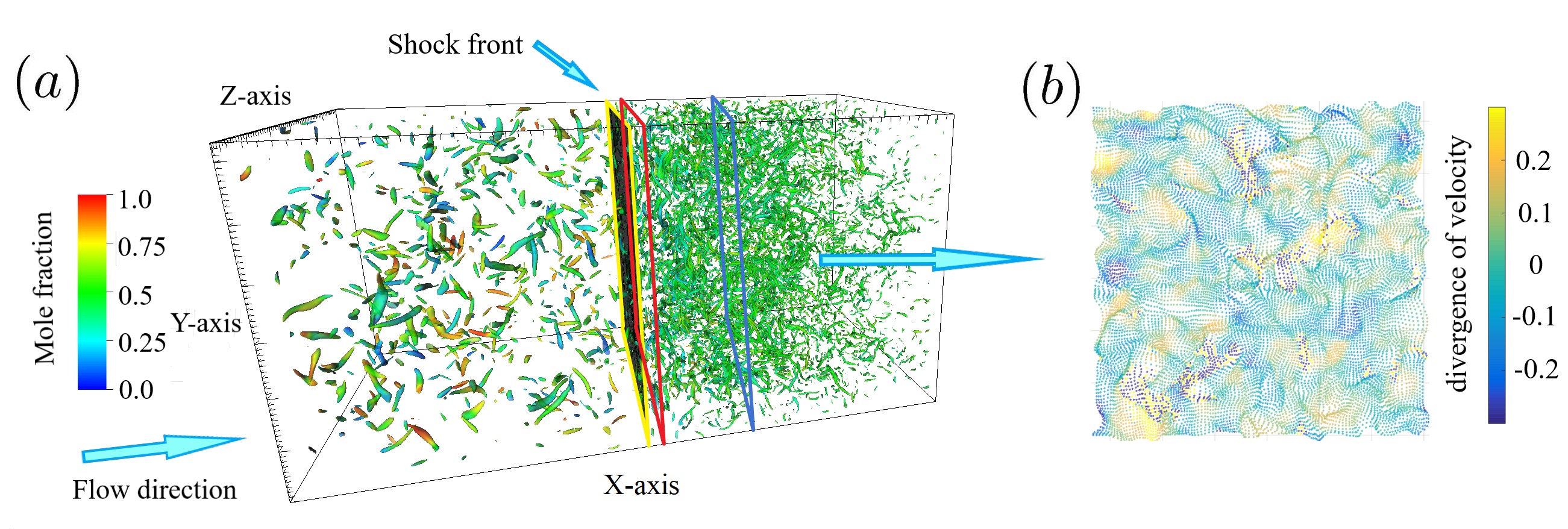}
		\caption{Instantaneous contours of vorticity and shock surface in isotropic turbulence interacting with a Mach 2 shock. (\textit{a}) Vortex structures are identified by the Q criterion (i.e. isosurface of the second invariant of VGT: $Q=2\left \langle Q_w \right \rangle$, where $\left \langle Q_w \right \rangle$ is the averaged magnitude of rotation tensor), colored by the mole fraction of the heavy fluid. Fluid particles are initialized as a sheet that spans over the homogeneous directions at a given post-shock streamwise position and allowed to develop with the flow. (\textit{b}) Visualized particle sheet, convected and distorted by the post-shock turbulence.}
		\label{fig:figure1}
	\end{figure*}
	
	\subsection{Numerical Setup}
	\label{subsec:setup}
	
	The physical domain for the simulations considered in this paper is a box that has a dimension of 4$\pi$ in the streamwise direction (denoted as $x$) and $(2\pi,2\pi)$ in the transverse directions (denoted as $y$ and $z$), as shown in figure \ref{fig:figure1} (a). The flow in this figure is visualized using the iso-surface of $Q$, the second invariant of the velocity gradient tensor $A_{ij}=\partial u_i / \partial x_j$. The normal shock is located at $x=2\pi$. A buffer layer is used at the end of the computational domain from $4\pi$ to $6\pi$ to eliminate reflecting waves. In the transverse directions, periodic boundary conditions are used as the flow is assumed to be periodic and homogeneous in these directions. To provide inflow turbulence, pre-generated decaying isotropic turbulence is superposed on the uniform mean flow with Mach number = 2.0 and convected into the domain using Taylor's hypothesis. The inflow turbulent Mach number, Reynolds number and peak wavenumber are $M_t \approx 0.1$, $Re_\lambda \approx 45$, and  $k_0=4$, respectively. For this $M_t$ value, Taylor's hypothesis is appropriate for approximating spatially developing turbulence with temporally developing turbulence \citep{lee1992simulation}. The variable density (multi-fluid) effects arise from compositional variations of a binary mixture of miscible fluids with different molar masses, which is generated by correlating the density to an isotropic scalar field representing the mole fraction of the heavy fluid. The scalar field is generated as a random field following a Gaussian spectrum with a peak at $k_s=8.0$ and has double-delta probability density function (PDF) distribution so that the scalar value initially is either 1.0 or 0.0. The initial scalar field is smoothed by solving a diffusion equation so that the scalar field can be fully resolved by the chosen mesh. The resulting scalar field is then allowed to decay in the fully developed isotropic turbulence setup for one eddy turn over time as a passive scalar. The density field is then calculated by imposing $X=\phi$ (where $X$ is the mole fraction of the heavy fluid). The generated variable density isotropic turbulence is then superposed onto the mean flow and allowed to develop into a more realistic state before reaching the shock wave. The Atwood number, $A_{t}=(MW_{2}-MW_{1})/(MW_{2}+MW_{1})$, calculated from the molar weights of the two fluids, $MW_{1}$ and $MW_2$, is 0.28. This value of the Atwood number was chosen such that the variable density effects are non-negligible, yet the interaction with the shock wave is still in the wrinkled-shock regime.  At larger Atwood numbers, the interaction enters the broken shock regime, where more complicated dynamics exist. The extension of the current study to this regime poses significant challenges, which are beyond the goals of the current study. The Prandtl number, $\Pran$, and Schmidt number, $\Sc$,  are the same and equal to 0.75. Immediately before the shock wave, $M_t$ and $Re_\lambda$ reach around 0.09 and 42 due to turbulence decay. For these values, the nonlinear and viscous effects on turbulence passing through the shock wave are weak based on the results of LIA convergence tests done in our previous study \citet{tian2017}.

	\subsection{Interpolation Scheme for the Lagrangian study}
	\label{subsec:scheme}
	
	For the current study, we have tracked more than 4.5 million particles that are initialized uniformly at various streamwise positions $\vec{x_{0}}$, and calculated various turbulence statistics following their trajectories. The aim is to understand the evolution of flow structures following fluid particles as the turbulence develops downstream of the shock. Figure \ref{fig:figure1} (a) marks with red lines a typical streamwise plane where particles are initialized. Each set of particles is initialized uniformly in the spanwise directions at the same streamwise location, corresponding to a planar sheet. The spacing between the neighboring particles in the spanwise directions is the same as the grid size (2$\pi$/512). We have uniformly sampled around 20 particle sets (sheets) for each cycle of the inflow turbulence box. The particles are then convected by the instantaneous turbulent velocity obtained by turbulence-resolving shock-capturing simulations and moved to a region marked by the blue lines. At this stage, the initially flat particle sheet is distorted by the turbulence as shown in figure \ref{fig:figure1} (b).
	
	The fluid particles are non-inertial and follow the local flow velocity. The corresponding transport equations for particle positions $x^{+}_{i}$ are:
	
	\begin{subeqnarray}
		\frac{dx^{+}_{i}(t|\vec{x_{0}},t_{0})}{dt} &= &u_{i}^{+}(t|\vec{x_{0}},t_{0}),\\
		u_{i}^{+}(t|\vec{x_{0}},t_{0}) &= &u_{i}(x^{+}_{i},t)
	\end{subeqnarray}

\noindent	
	where $x^{+}_{i}(t|\vec{x_{0}},t_{0})$ represents the positions of the particles at time $t$ that are initialized at $\vec{x_{0}}$ and time $t_{0}$. The particle velocity $u_{i}^{+}(t|\vec{x_{0}},t_{0})$ can be obtained from the Eulerian velocity field $u_{i}(x^{+}_{i},t)$ by interpolation. The interpolation is based on the cubic spline scheme, whose accuracy in predicting particle positions has been studied in \citet{yeung1988algorithm}. The time-stepping scheme for Lagrangian particles is also the third-order Runge-Kutta scheme. Therefore, at each sub-timestep, the particle velocity is interpolated from the Eulerian velocity field with the same sub-timestep. In the STI configuration, there is a sharp change of the flow velocity at the shock, which deteriorates the interpolation accuracy. To achieve accurate interpolation of the particle velocity, the domain is partitioned into three different regions as shown in figure \ref{fig:figure1} (a): pre-shock, shock, and post-shock regions. The instantaneous shock surface is identified using the sensor: $s=-\theta/(|\theta|+\langle \omega_i\omega_i \rangle_{yz}^{0.5})>0.5$ \citep{larsson2009}, where $\theta=\partial u_i/\partial x_i$ is the dilatation, $\omega_i=\epsilon_{ijk}\partial u_k/\partial x_j$ is the vorticity, and $\langle \rangle_{yz}$ represents the instantaneous average over the homogeneous directions. After the instantaneous shock region is identified, the pre- and post-shock turbulence fields can be separated for interpolation. Note that the cubic spline interpolation scheme requires information from neighboring cells, so a buffer region (around three grid points) is added between the shock region and the post-shock region. Lagrangian dynamics of particles across the shock wave is not considered in this study because the shock profile is numerical and its thickness depends on the grid size. This introduces numerical artifacts when considering the particle dynamics across the shock wave.

	\subsection{Grid and Statistical Convergence}
	\label{subsec:converge}
	
\begin{figure}
		\centering
		\includegraphics[width=5in]{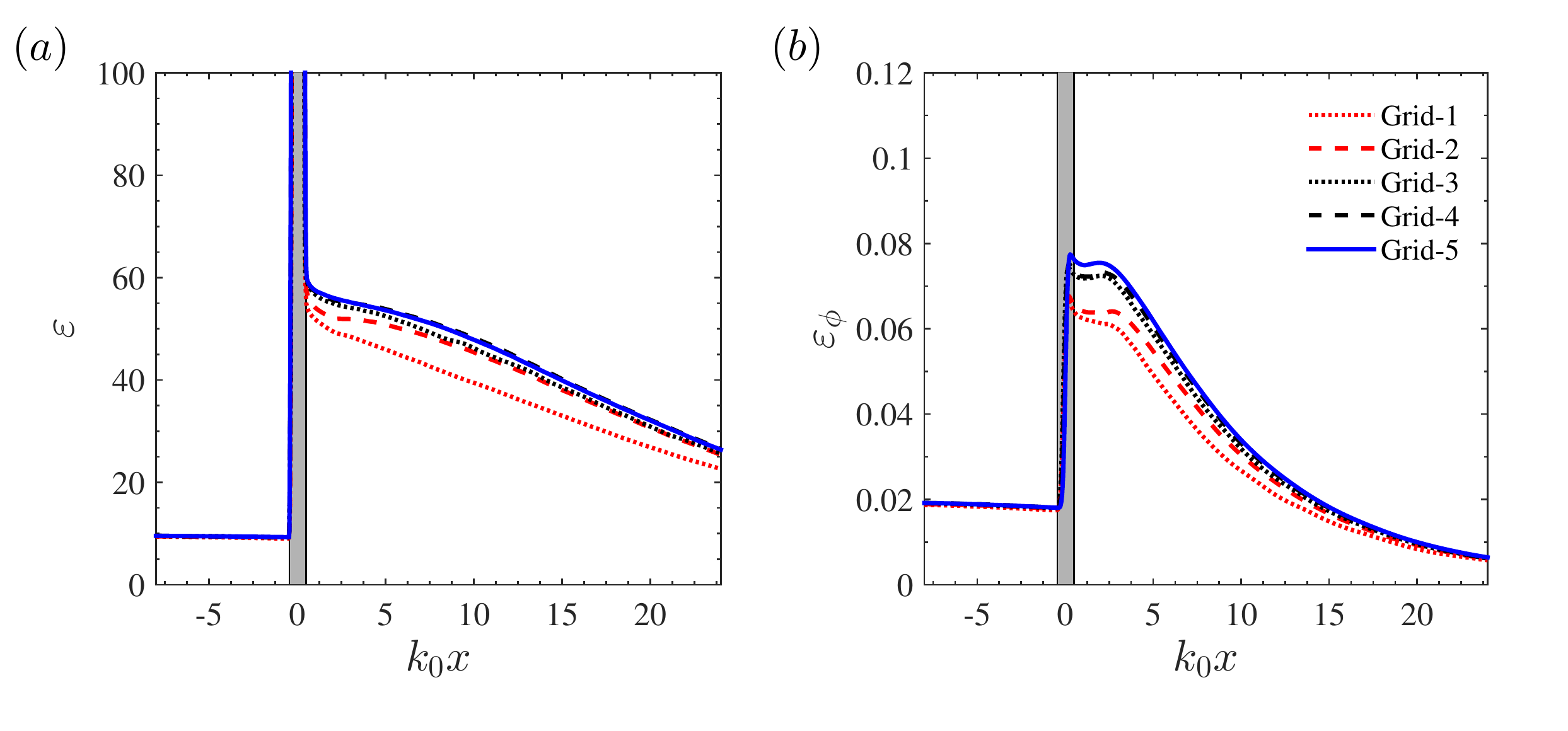}
		\caption{Results of multi-fluid grid convergence tests at $Re_{\lambda}=45$ and $M_{t}=0.1$. Streamwise development of (\textit{a}) turbulent dissipation rate $\varepsilon$ and (\textit{b}) mass fraction dissipation rate $\varepsilon_\phi$ is shown. The region of unsteady shock movement is marked in grey.}
		\label{fig:converg}
	\end{figure}

	The accuracy of the numerical results is addressed in this subsection through a series of convergence tests. To ensure that all the turbulence length scales are well resolved, a grid convergence test was conducted in \citet{tian2017}. Here, we summarize these results for completeness, together with additional convergence results for small-scale quantities. Figure \ref{fig:converg} shows the turbulence dissipation rate $\varepsilon=-\overline{\sigma_{ij} \frac{\partial u_{i}}{\partial x_{j}}}$, where $\sigma_{ij}$ is the viscous stress tensor, and scalar (mass fraction for the multi-fluid STI) dissipation rate $\varepsilon_\phi=\overline{\frac{\mu}{\Rey_{0}\Sc} \frac{\partial\phi}{\partial x_{j}}\frac{\partial \phi}{\partial x_{j}}}$ as a function of the normalized streamwise direction $k_{0}x$ for a series of meshes. The grey regions in the following figures indicate the unsteady shock region, inside which the results are affected by the shock wrinkling and unsteady shock movement. As the grid is refined in all three directions, both quantities display convergence, proving the accuracy of the turbulence database. Another issue that needs to be considered is the scale separation between the numerical shock thickness $\delta_n$ and the Kolmogorov length scale $\eta$ as suggested in our previous study \citep{tian2017}. $\delta_{n}$ is calculated as $(u_{1,u}-u_{1,d})/ |\partial u_{1}/\partial x_{1}|_{max}$, and $|\partial u_{1}/\partial x_{1}|_{max}$ denotes the maximum magnitude of streamwise velocity gradient. Grid numbers for Grid 1 to 5 shown in figure \ref{fig:converg} are 256$\times$256$\times$1024, 384$\times$384$\times$1024, 384$\times$384$\times$1536, 512$\times$512$\times$1536, 512$\times$512$\times$2048. With the finest mesh (512$\times$512$\times$2048), the scale separation ratio $\eta/\delta_n$ is around 1.9, which is sufficient for resolving the interaction between the numerical shock wave and small-scale turbulent motions. Therefore, in the current study, we have obtained all the statistics from the turbulence field based on the finest grid to ensure accuracy. Finally, LIA convergence tests were conducted in \citet{tian2017} following \citet{ryu2014} to show that the shock-capturing simulations can capture the correct limits. Turbulent Mach number $M_t$ and Taylor Reynolds number $Re_\lambda$ were varied for the canonical single-fluid simulations, covering a wide range of parameter space. The shock-capturing simulation results do converge to LIA predictions for individual Reynolds stress components as long as certain conditions are satisfied \citep{tian2017}. This was the first time that the asymptotic values for individual Reynolds stresses were approximated using shock-capturing simulations. 
	
\begin{figure}
		\centering
		\includegraphics[width=5in,height=2in]{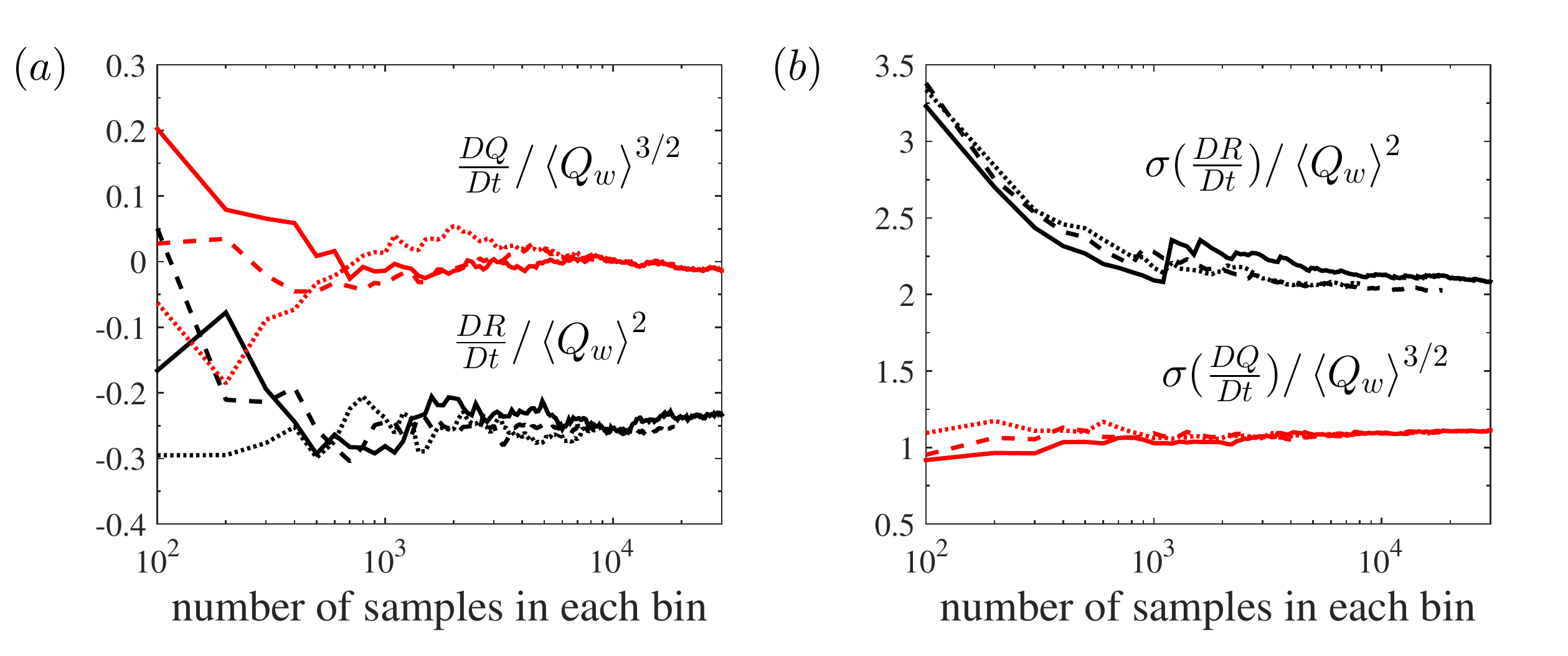}
		\caption{The statistical convergence for (\textit{a}) $(DQ/Dt)/\left \langle Q_w \right \rangle^{3/2}$ and $(DR/Dt)/\left \langle Q_w \right \rangle^{2}$ and (\textit{b}) their standard deviations conditioned at point (3.0,3.0) in the $(Q, R)$ phase plane for multi-fluid case.}
		\label{fig:staconverg}
	\end{figure}

\begin{figure}
		\centering
		\includegraphics[width=5in,height=2in]{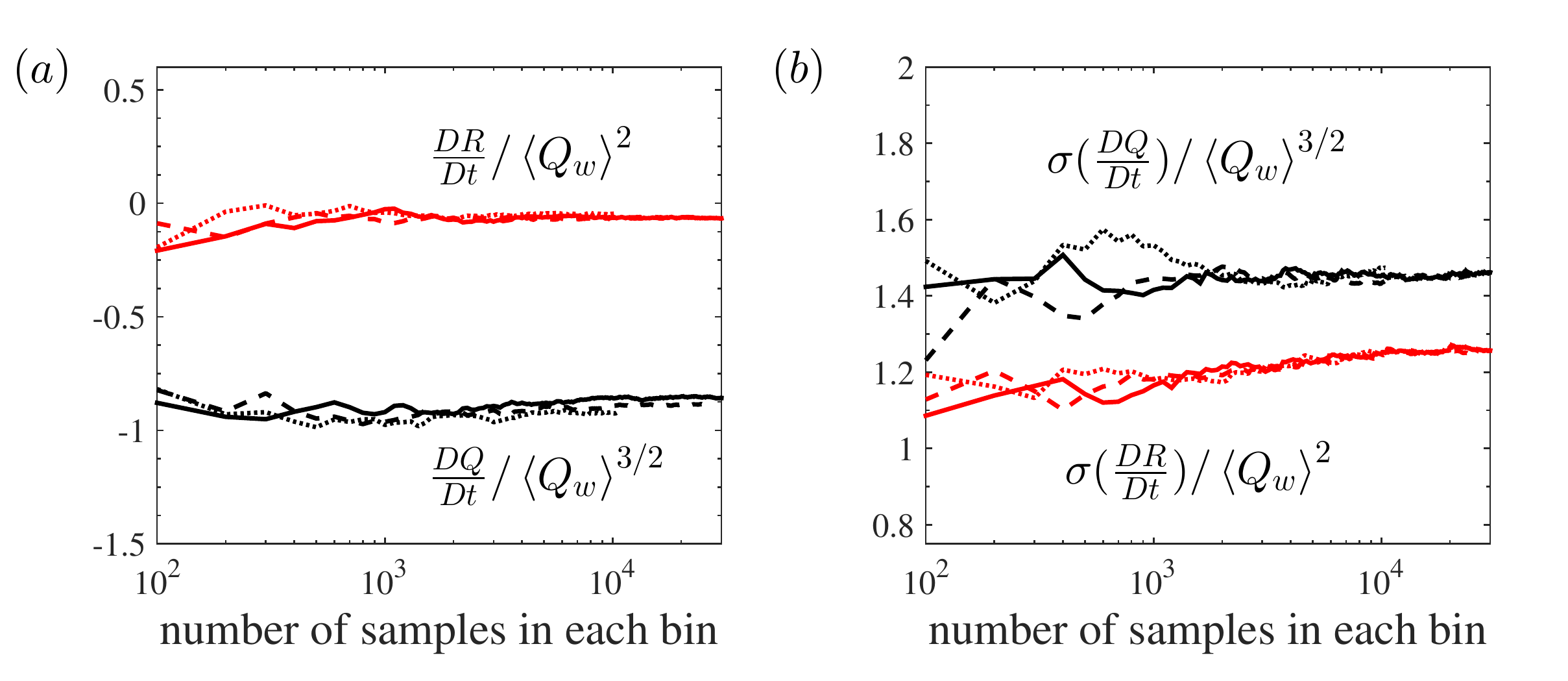}
		\caption{The statistical convergence for (\textit{a}) $(DQ/Dt)/\left \langle Q_w \right \rangle^{3/2}$ and $(DR/Dt)/\left \langle Q_w \right \rangle^{2}$ and (\textit{b}) their standard deviations conditioned at point (3.0,3.0) in the $(Q, R)$ phase plane for single-fluid case.}
		\label{fig:staconvergs}
	\end{figure}

	Statistical convergence is another important factor that needs to be addressed. To reduce the statistical variability, all the results that are based on the Eulerian data are space-averaged over homogeneous directions and time-averaged for around two flow-through times. The averaging is performed after the flow has reached a statistically steady-state to eliminate the effects of transient processes \citep{larsson2013}. For the Lagrangian statistics, the number of fluid particles needs to be large enough for statistical convergence, especially for conditional averaged statistics. The conditional averaged value of $X$, conditioned on the variable $A$ and $B$, is defined as:
	
    \begin{eqnarray}
		\left \langle X|(A=A_0,B=B_0) \right \rangle 
		 &=& \langle X|(A_0-\frac{1}{2}\Delta A) \leq A < (A_0+\frac{1}{2}\Delta A), \nonumber \\
		&&(B_0-\frac{1}{2}\Delta B) \leq B < (B_0+\frac{1}{2}\Delta B) \rangle
    \label{eqn:condition}
	\end{eqnarray}
where $\Delta A$ and $\Delta B$ are the bin sizes. The conditional statistics are obtained by ensemble averaging (denoted by $\langle \rangle$) over all the fluid particles that fall into the bins. Figure \ref{fig:staconverg} and \ref{fig:staconvergs} show the convergence of two important conditional Lagrangian statistics $\langle \frac{DQ}{Dt}\rangle/\left \langle Q_w \right \rangle^{3/2}$, $\langle \frac{DR}{Dt} \rangle/\left \langle Q_w \right \rangle^{2}$ and their standard deviation, depending on the number of particles in each bin. Here, $\frac{DQ}{Dt}$ and $\frac{DR}{Dt}$ represent the material derivative of the second invariant ($Q$) and third invariant ($R$) of the VGT. For the multi-fluid case, we note that the convergence of both conditional means and standard deviations can be achieved when using around 10,000 particles, larger than that needed for the canonical single-fluid case as shown in figure \ref{fig:staconvergs}. This suggests that the variable density effects make the simulations more computationally demanding. The effects of the bin sizes are also examined by comparing three different set of bin numbers $30\times30$ (solid), $40\times40$ (dashed) and $60\times60$ (dotted) in the $(Q, R)$ phase plane at the same point (3.0,3.0). These bin numbers correspond to the following bin sizes: $(1.3,1.3)$, $(1.0,1.0)$ and $(0.67,0.67)$. Our analysis indicate that the statistics converge to almost the same values when the sample size is large enough. In the present study, we uniformly sampled more than 4.5 million particles and made sure that there are at least 10,000 particles in each sample bin with the number of bins being $40\times40$ ($(\Delta Q,\Delta R)=(1.0,1.0)$). 

	\section{Results and Discussions}
	\label{sec:results}
	
	The variable density effects on the post-shock turbulence structure and dynamics are examined in this section. The results obtained from the multi-fluid STI simulation are compared with those of a reference single-fluid case and standard isotropic turbulence. First, the post-shock turbulence state and its evolution away from the shock wave are examined to identify the variable density effects. The results are based on time- and space-averaged statistics obtained from the Eulerian data. The flow topology is studied next to further understand the post-shock turbulence evolution.  The dynamics that dominate the transient evolution of post-shock turbulence structure are examined using the Lagrangian equation of VGT and Lagrangian data collected for sample fluid particles.

	\subsection{Density effects on post-shock turbulence}
	
	\label{subsec:deneffects}
	\subsubsection{Turbulence state immediately after the shock}
	
	\label{subsubsec:turbstate}
	In this section, the turbulence structure immediately after the shock wave is analyzed to identify the different roles that density plays through the shock wave.
	
\begin{figure}
		\centering
		\includegraphics[width=5in]{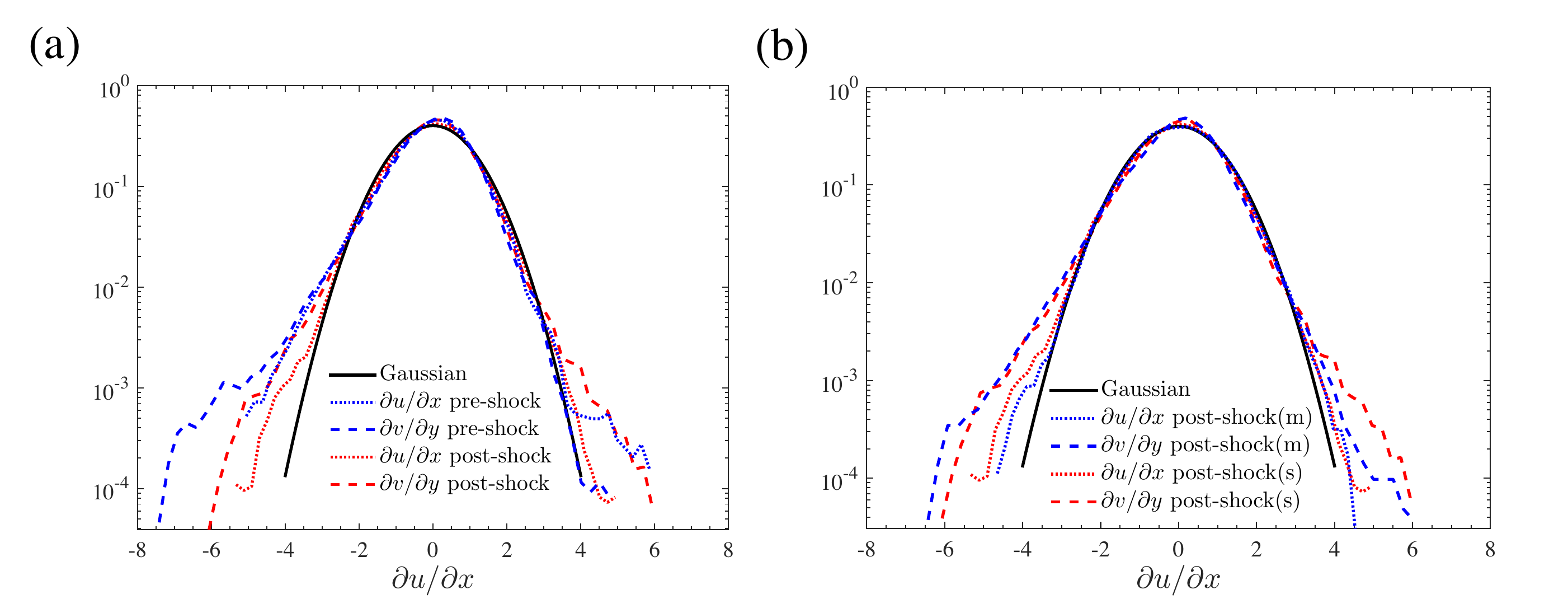}
		\caption{Comparison of the PDFs of the normalized post-shock velocity derivatives with a Gaussian distribution. Comparison of (\textit{a}) multi-fluid pre-shock with post-shock results and (\textit{b}) multi-fluid with single-fluid post-shock results.}
		\label{fig:vgpdf}
	\end{figure}

	The PDFs of streamwise and spanwise longitudinal velocity derivatives for pre- and post-shock ($k_0x=0.5$) multi-fluid turbulence are shown in figure \ref{fig:vgpdf} (a) alongside the Gaussian distribution as a reference. The non-Gaussian nature of the velocity gradient PDFs and their connection to the energy cascade and intermittency are well documented in the turbulence literature. The PDFs of the pre-shock velocity derivatives are negatively skewed as expected. After passing the shock wave, they become closer to the Gaussian distribution, especially for the streamwise component. The PDFs for both single-fluid and multi-fluid post-shock turbulence are shown in figure \ref{fig:vgpdf} (b).  Here, we note that immediately after the shock wave, the PDF of the spanwise velocity gradient for both cases remains negatively skewed, as in isotropic turbulence. The streamwise component, however, becomes more symmetric and Gaussian-like due to the interaction with the shock wave. This indicates that the energy transfer to small scales is suppressed in the streamwise direction. We also note that the density has a relatively weak effect on the velocity derivatives PDFs since the single-fluid and multi-fluid cases have similar PDFs.
	
\begin{figure}
		\centering
		\includegraphics[width=3in]{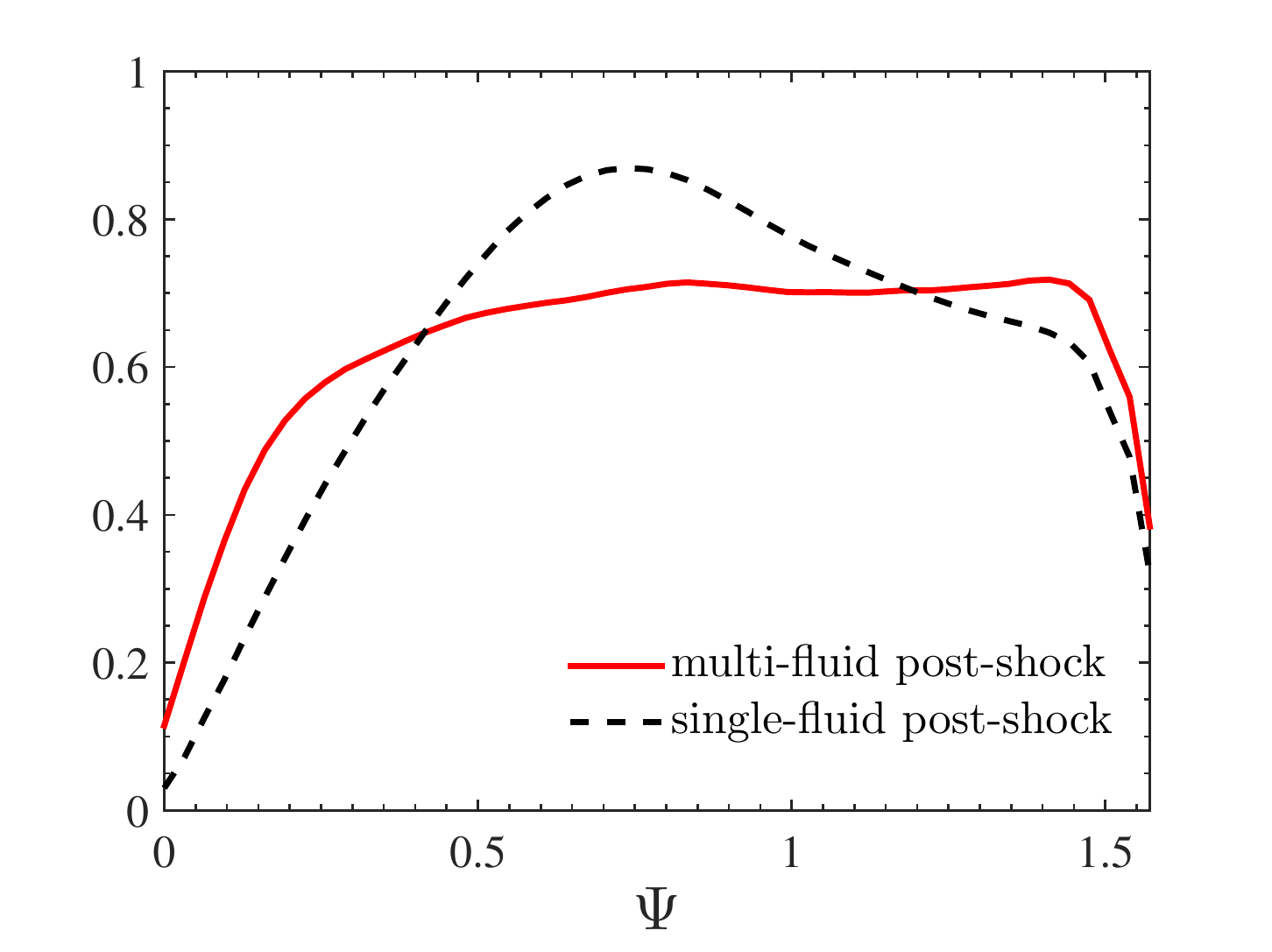}
		\caption{PDF of the strain-enstrophy angle $\Psi$ in radians for post-shock turbulence.}
		\label{fig:pdf_ens}
	\end{figure}

	The preferential amplification of the transverse components of the rotation and strain rate tensors is an important effect in STI and has been extensively studied for the canonical single-fluid flows \citep{mahesh1997,ryu2014,livescu2015vorticity}. This amplification can lead to an increase in the correlation between the two quantities. To better understand the variable density effects on post-shock turbulence, the PDF of the strain-enstrophy angle, $\Psi$, is considered in figure \ref{fig:pdf_ens}. $\Psi$ is calculated using $\Psi=$tan$^{-1} (S_{ij}S_{ij}/(W_{ij}W_{ij}))$, where $S_{ij}=1/2(A_{ij}+A_{ji})$ and $W_{ij}=1/2(A_{ij}-A_{ji})$ are the strain and rotation tensors.  In isotropic turbulence, the PDF of $\Psi$ peaks near $90^{\circ}$ \citep{jaberi2000characteristics}, indicating a strain dominated flow. In single-fluid post-shock turbulence, the PDF of $\Psi$ exhibits a shift of the peak from $90^{\circ}$ to around $45^{\circ}$, as the shock Mach number increases. This has been observed by \citet{livescu2015vorticity} and is interpreted as the increase in correlation of strain and rotation. However, in the multi-fluid case, the peak still occurs at relatively large angles and the increase in correlation is not as pronounced as that in the single-fluid case, at the same shock Mach number. Figure \ref{fig:pdf_ens} implies that the rotation and strain are amplified differently by the shock when large density variations are present, which compromises the correlation between the two quantities. 
	
	\begin{figure}
		\centering
		\includegraphics[width=3in]{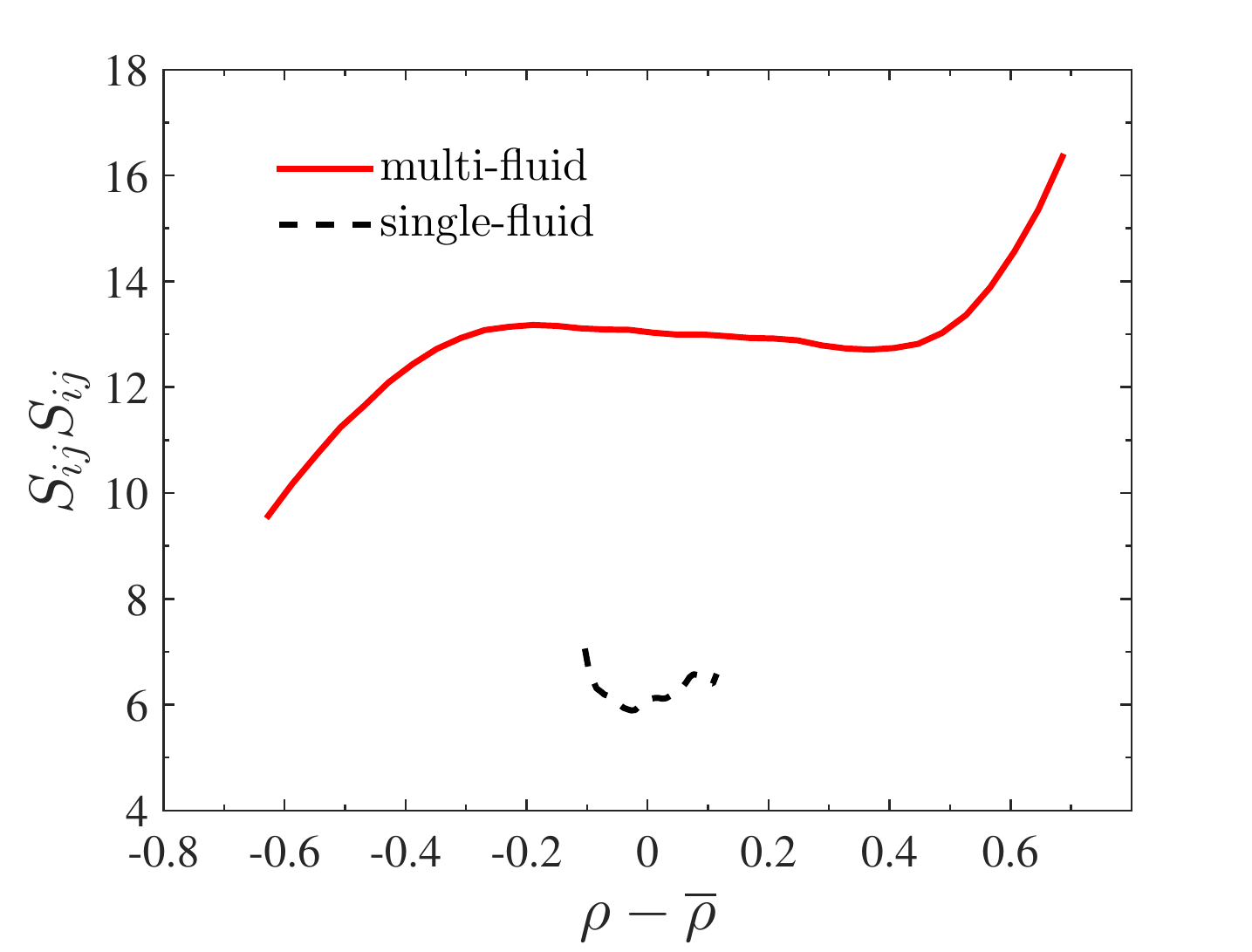}
		\caption{Conditional expectation of the magnitude of strain rate tensor as a function of density after the shock wave.}
		\label{fig:dencon}
	\end{figure}
	
\begin{figure}
		\centering
		\includegraphics[width=3in]{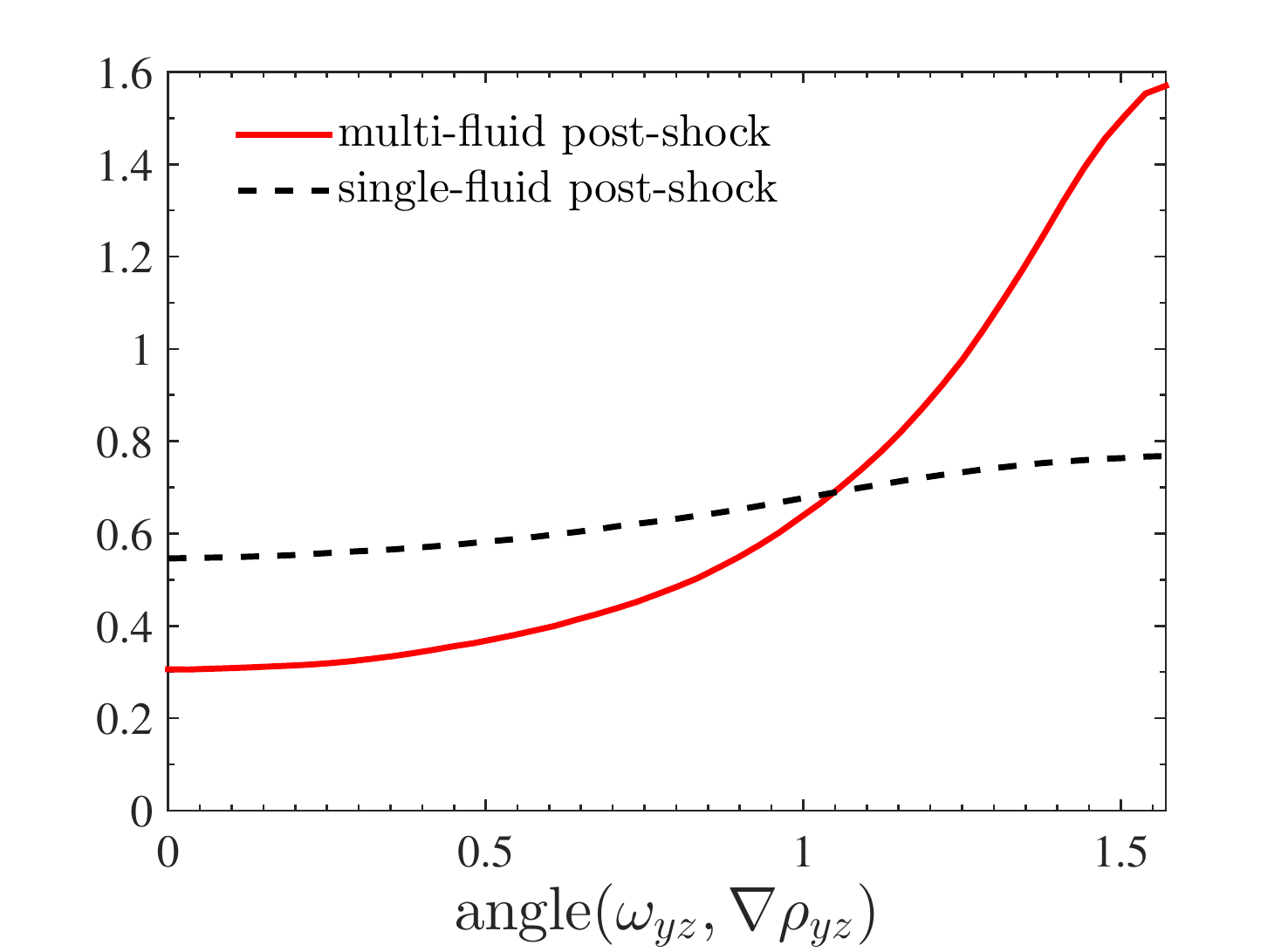}
		\caption{PDF of the orientation between the vorticity vector and density gradient in y-z direction immediately after the shock wave.}
		\label{fig:baro}
	\end{figure}

	The variable density effects on strain and rotation tensors can be studied by examining the conditional expectations of their magnitudes as a function of density. It was shown in \citet{tian2017} that through the shock wave, the amplification of vorticity is stronger in the mixed fluid regions with near-average density, but weaker in the pure fluid regions. This is not observed in the single-fluid simulation. One mechanism that might be responsible for this behavior is the baroclinic torque: $(\nabla \rho \times \nabla p)/\rho^2$ in the vorticity transport equation. A strong pressure gradient $\nabla p$ exists through the shock wave; at the same time, large density gradients $\nabla \rho$ also exist, especially in the mixed fluid regions. Since the pre-shock density field is isotropic, $\nabla \rho$ and $\nabla p$ can be locally misaligned, especially when the spanwise component of $\nabla \rho$ is large, becoming a source of vorticity generation through the baroclinic torque. In addition, the generated vorticity field should be perpendicular to the spanwise density gradient. In the pure fluid regions or single-fluid simulation, however, the density gradients are much smaller, so that the cross product of $\nabla p$ and $\nabla \rho$ is also small. Note that the density gradient in the streamwise direction has no contribution, because it is aligned with the pressure gradient. To confirm this, the PDF of the angle between the spanwise component of density gradient and the vorticity vector is plotted in figure \ref{fig:baro}. After the shock wave, the multi-fluid case exhibits a stronger tendency of the vorticity vector being perpendicular to the density gradient. In contrast, this tendency is not observed in the single-fluid case. This provides evidence that density gradient and baroclinic torque play important roles in establishing the preferential deposition of vorticity across the shock wave.
	
\begin{figure}
		\centering
		\includegraphics[width=5in]{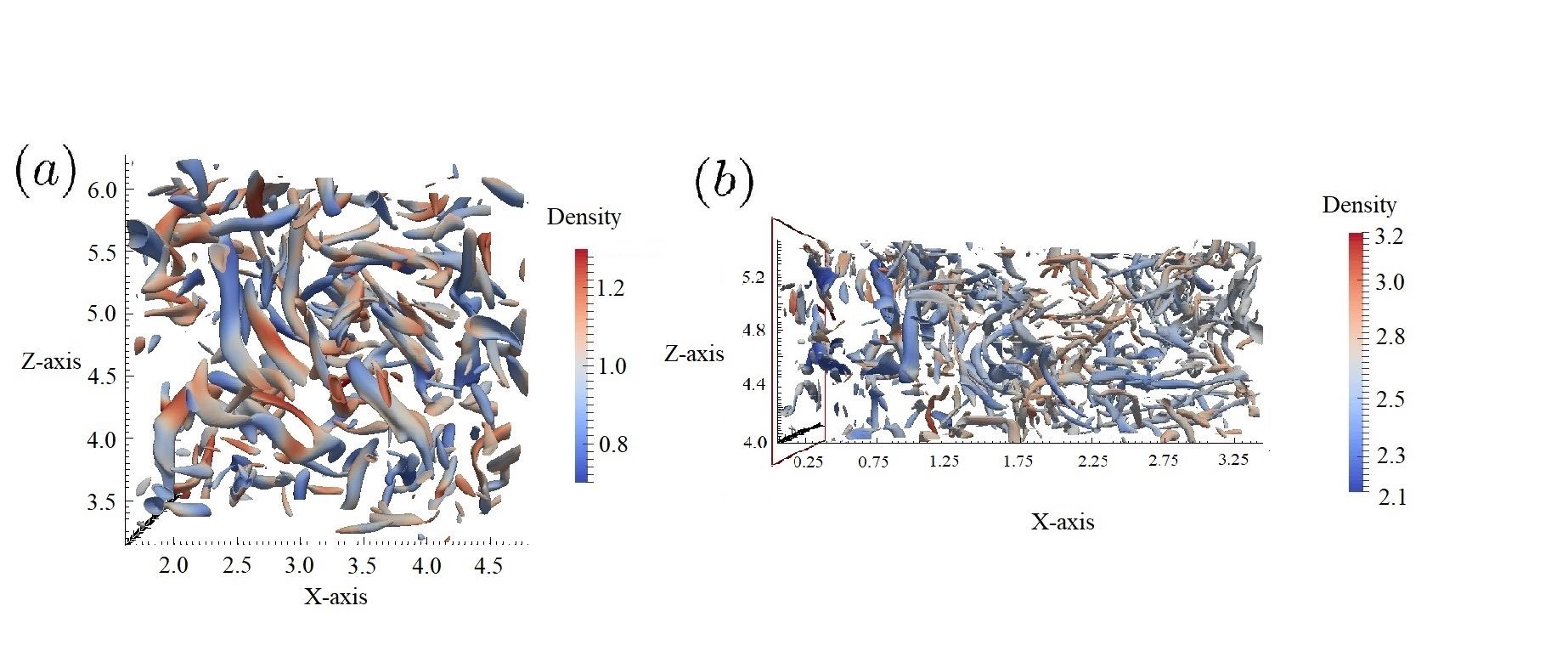}
		\caption{Vortex structures captured using the Q-criterion, colored by density, for multi-fluid (\textit{a}) pre-shock turbulence and (\textit{b}) post-shock turbulence.}
		\label{fig:vorden}
	\end{figure}

	Figure \ref{fig:vorden} can help visualize the changes in the flow structure across the shock wave. The vortex tubes are captured using the Q-criterion and are colored by their local density. Figure \ref{fig:vorden} (a) shows the vortex structures for pre-shock multi-fluid isotropic turbulence. For the visualized vortex tubes, there are no identifiable effects from the density variations; the vortex tubes are not preferentially distributed due to the density effects. However, the interaction with the shock has a clear effect on the post-shock vortical structures (figure \ref{fig:vorden} b). Immediately behind the shock wave, vortex tubes are aligned in the spanwise direction, which has been observed in previous STI studies \citep{larsson2013,boukharfane2018evolution}. More importantly, the vortex tubes also get aligned with the density iso-surfaces, meaning that the vorticity becomes perpendicular to the density gradient. This is consistent with the earlier analysis of the baroclinic torque. As a consequence, the post-shock vorticity field enhances the mixing between adjacent density regions. This coupling is further explored in the next section.

	For the strain rate tensor, figure \ref{fig:dencon} shows that its magnitude tends to be stronger in the heavy fluid regions and weaker in the light fluid region. This trend is hypothesized to be related to the dependence of shock strength on the pre-shock density. \citet{tian2017} showed that shock compression is stronger in the heavy fluid region, while it is weaker in the smallest density regions, leading to the observed trend in the amplification of the magnitude of the strain rate tensor. This trend is different from that observed for the vorticity, which is explained above. As a result, the trend of the strain-enstrophy angle PDF peaking around $45^{\circ}$, observed in the single-fluid case at higher shock Mach numbers, is weakened in the multi-fluid case. Identifying the specific mechanisms behind variable density turbulence interactions with shock wave, such as shock intensity dependence on density, density gradient effects, inertial effects and so on, can potentially be beneficial to modeling variable density STI.
	
	\subsubsection{Evolution of turbulence state downstream of the shock}
	
	The evolution of variable density turbulence away from the shock wave involves many coupled nonlinear processes. In this section, the focus is on the evolution of turbulence structures. 
	
\begin{figure}
		\centering
		\includegraphics[width=5in]{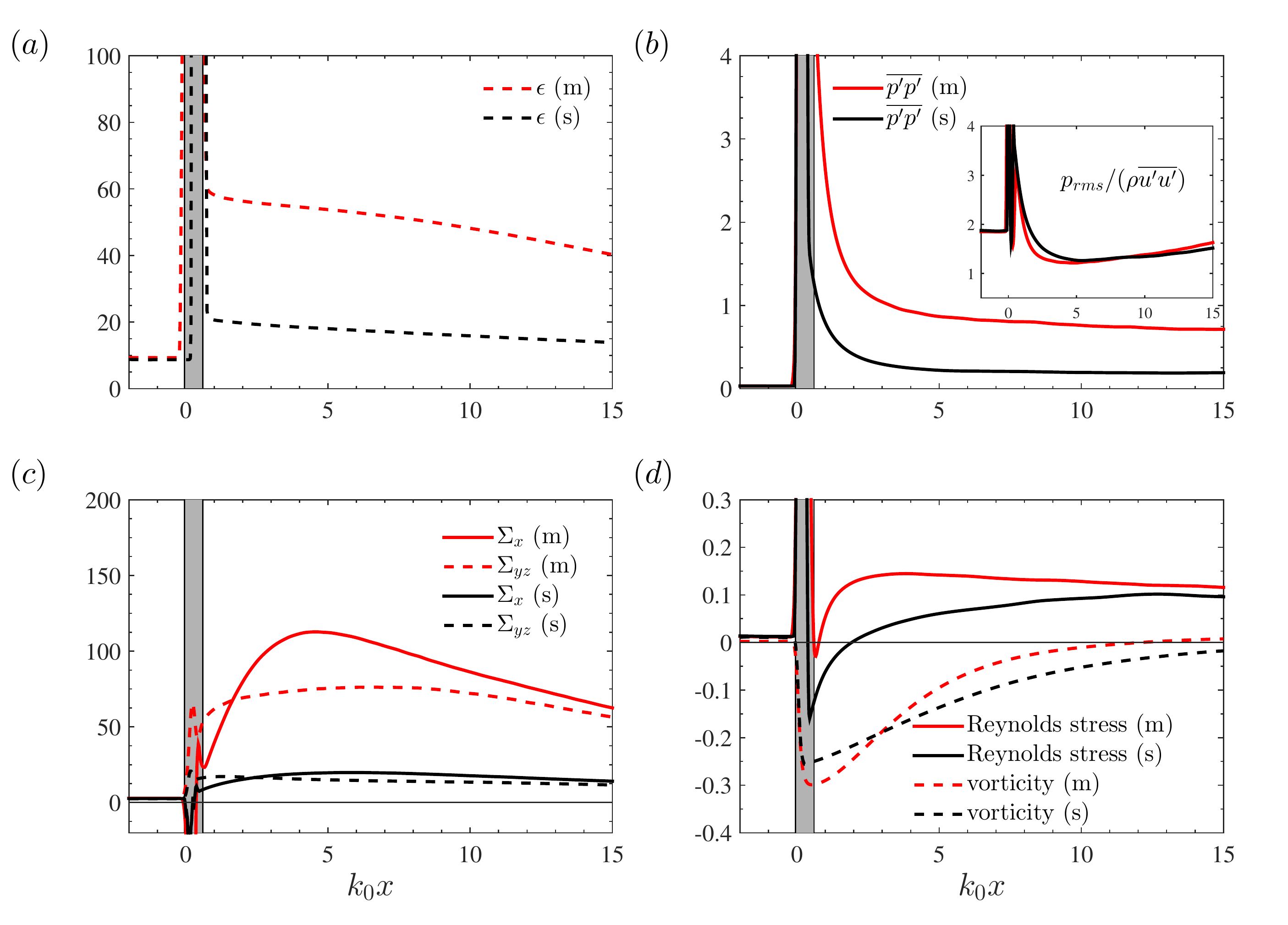}
		\caption{Development of \textit{(a)} turbulence dissipation rate, \textit{(b)} pressure variance, \textit{(c)} vortex stretching, and \textit{(d)} anisotropy ($b_{11}$) of Reynolds stress and vorticity.}
		\label{fig:compare}
	\end{figure}

	Figure \ref{fig:compare} shows the development of some of the fundamental turbulence statistics. The evolution of these statistics helps in the understanding of the general characteristics of single- and multi-fluid STI. Figure \ref{fig:compare} (a) shows that with the introduction of strong density variations, the shock amplification of dissipation rate is stronger. Figure \ref{fig:compare} (b) shows the fluctuating pressure variance as a function of the streamwise position to highlight the development of the acoustic field. The amplification of the pressure fluctuations across the shock wave is noted, agreeing with \citet{sethuraman2018thermodynamic}. The acoustic wave is stronger in the multi-fluid case immediately after the shock wave. This is related to the shock intensity fluctuations induced by the strong density variations. As a result, the decay of the acoustic field is also faster for the multi-fluid case, causing a faster increase in TKE. After the post-shock transient pressure adjustment, the multi-fluid case still exhibits larger absolute pressure fluctuations. However, after normalizing with $\rho\overline{u'u'}$, the pressure fluctuations become somewhat similar in magnitude in these two cases. In figure \ref{fig:compare} (c), the vortex stretching term $\Sigma=\overline{\omega_{i}\omega_{j}\frac{\partial u_{i}}{\partial x_{j}}}$ is decomposed into its streamwise $\Sigma_x=\overline{\omega_{1}\omega_{j}\frac{\partial u_{1}}{\partial x_{j}}}$ and spanwise $\Sigma_{yz}=\overline{\omega_{2}\omega_{j}\frac{\partial u_{2}}{\partial x_{j}}}$ components to explore the axisymmetric state and return-to-isotropy of post-shock turbulence. Previous studies \citep{livescu2015vorticity} have demonstrated that the normalized vortex stretching term reaches a low value after passing through the shock wave, indicating a tendency towards an axisymmetric state. Without normalization, Figure \ref{fig:compare} (c) shows that the absolute values of the vortex stretching terms are magnified in both single- and multi-fluid cases, more so for the spanwise component. The two components then undergo a transient process, where they first increase and cross each other, before the flow returns to an isotropic state. 
	
	In order to quantitatively study the evolution of  turbulence anisotropy, we consider here the Reynolds stress anisotropy tensor defined as $b_{ij}=\overline{u_i'u_j'}/\overline{u_k'u_k'}-\frac{\delta_{ij}}{3}$. A similar anisotropy tensor, $d{ij}$, can also be defined for the vorticity field, as $d_{ij}=\overline{\omega_i' \omega_j'}/\overline{\omega_k' \omega_k'}-\frac{\delta_{ij}}{3}$. Due to the homogeneity in spanwise directions, the diagonal components of the anisotropy tensor are related by $b_{22}=b_{33}=-0.5b_{11}$, so only $b_{11}$ is discussed. The near-zero value of $b_{11}\approx 0.0$ is an indication that flow has reached an isotropic state, while $b_{11}\approx -1/3$ means that the turbulent field has a tendency towards a 2D axisymmetric state. Figure \ref{fig:compare} (d) shows that $d_{11}$, a small-scale turbulent variable, attains value near -0.3 in the multi-fluid case, which is lower than that observed for the single-fluid case. This indicates that density intensifies the trend towards axisymmetric state for small-scale turbulence. On the other hand, the stronger turbulent stretching mechanism as observed in figure \ref{fig:compare} (c), makes the return to isotropy much faster in the multi-fluid case as compared to that in the single-fluid case. For Reynolds stresses, large-scale turbulent variables, the multi-fluid flow reaches a quasi-isotropic state immediately after the shock wave ($b_{11} \approx 0.0$), while single-fluid turbulence exhibits a tendency towards an axisymmetric state. This is in good agreement with \citet{boukharfane2018evolution}. Evidently, the variable density effects on the post-shock turbulence appear differently at small and large scales. Additionally, the quasi-isotropic state of the multi-fluid turbulence is not stable and is modified in the post-shock transition. Due to the energy transfer between the acoustic field and solenoidal turbulence field, $R_{11}$ quickly increases, causing $b_{11}$ to become larger than zero. The anisotropy reaches its maximum value around the peak TKE position ($k_0x \approx 2.0$) and then slowly decreases. For the single-fluid case, $b_{11}$ keeps increasing till $k_0x \approx 13.0$, even though the acoustic effects almost vanish after peak TKE location of $k_0x \approx \pi$.

\begin{figure}
		\centering
		\includegraphics[width=5in]{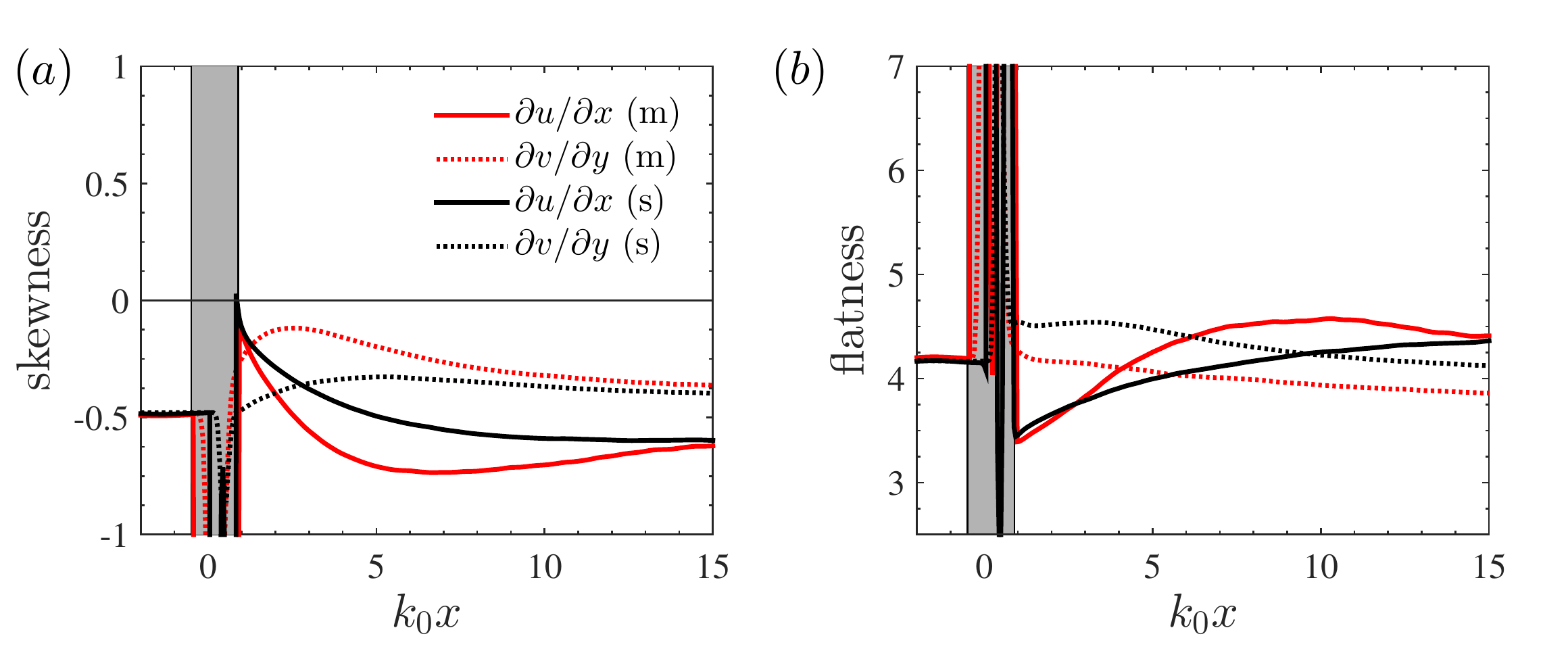}
		\caption{Development of \textit{(a)} skewness, and \textit{(b)} flatness, of the streamwise and transverse components of velocity derivatives.}
		\label{fig:skewflat}
	\end{figure}

	In figure \ref{fig:skewflat}, the developments of skewness and the flatness of the longitudinal velocity gradients are examined before and after the flow interaction with the shock wave. They show how the non-Gaussian behavior of the velocity field and specifically VGT are affected by the combined shock and density effects. For isotropic turbulence, the skewness of the longitudinal velocity gradient should be around -0.5, which is observed to be true in the pre-shock region for both single and multi-fluid cases for both streamwise as well as spanwise components (figure \ref{fig:skewflat} a). Immediately after the shock, different components of the derivative skewness tensor are shown to be modified in different ways. The streamwise component for both single-fluid and multi-fluid cases approaches to values very close to 0.0, which is consistent with the tendency towards a two-dimensional axisymmetric state observed above. As the turbulence evolves away from the shock wave, the streamwise velocity derivative skewness decreases rapidly. Due to the strong density variations, the multi-fluid case exhibits a faster decrease in skewness before $k_0x=5.0$, after which it slowly increases towards the $-0.54$ value. The shock modification of the skewness of the transverse derivative is relatively small for the single-fluid case. For the multi-fluid case, the longitudinal transverse velocity derivative becomes less negatively skewed, with a value of around -0.25. This difference can be attributed to stronger shock intensity variations and shock wrinkling in the multi-fluid case.  Away from the shock wave, for both cases, the skewness of $\partial v /\partial y$ increases first until it reaches a peak and then slowly decreases. Comparably, the multi-fluid case exhibits a shorter but more intensified transition. At the end of the domain, however, the spanwise derivative skewness is still larger than $-0.5$, as the flow is still anisotropic. Figure \ref{fig:skewflat} (b) shows the development of longitudinal velocity derivative flatness factor across and after the shock wave. Immediately after the shock, the flatness of the streamwise component decreases in value while that of the spanwise component increases. Similar to the skewness, the effect of density variations is relatively small on the flatness of streamwise component for the Atwood number considered in this study.  On the other hand, the density variations in the multi-fluid case make the increase in flatness of transverse component less significant, with the pre- and post-shock values being almost the same. Away from the shock wave, the flatness of the longitudinal streamwise velocity derivative increases, returning to its pre-shock value, while the growth is much faster in the multi-fluid case. For the transverse longitudinal derivative component, the flatness slowly decreases after a small change.

	\begin{figure}
		\centering
		\includegraphics[width=5in]{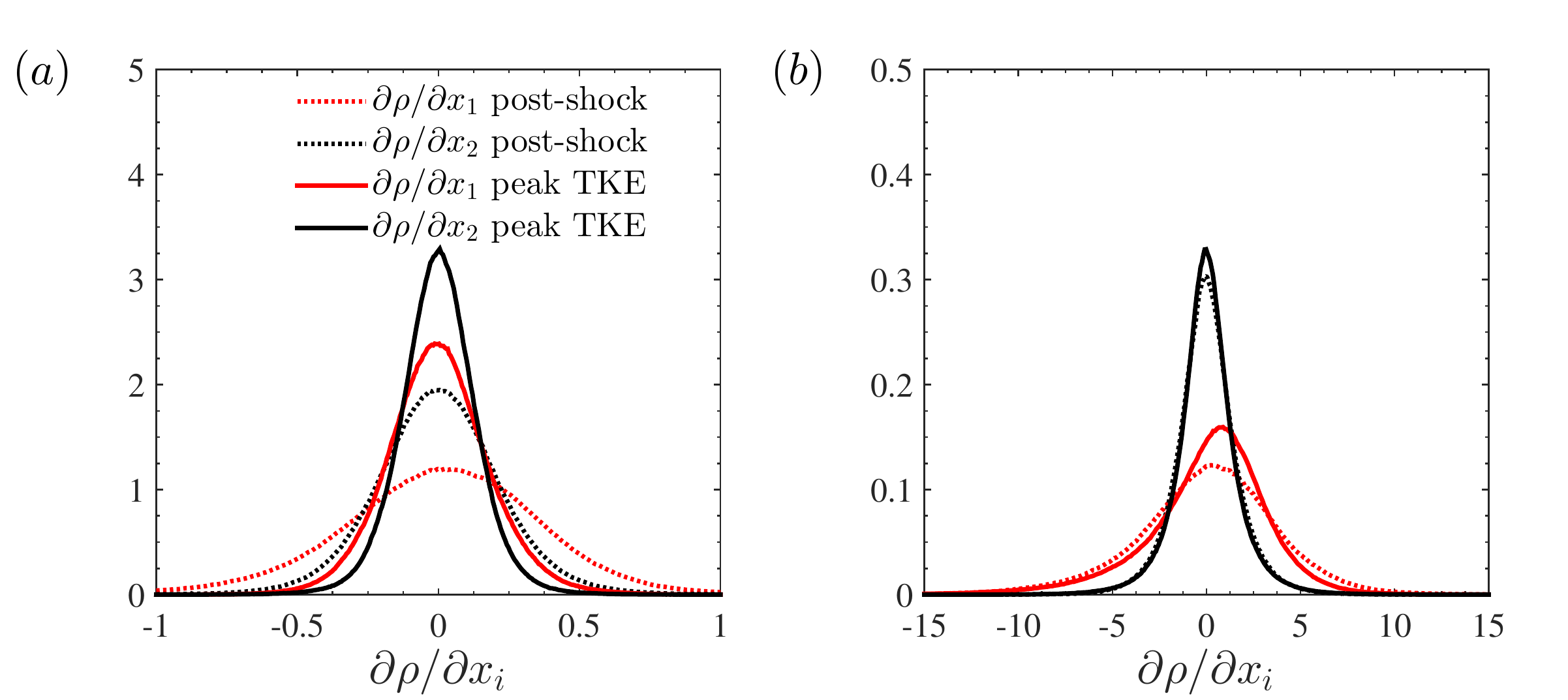}
		\caption{PDF of the density gradient at different streamwise locations for: (\textit{a}) single-fluid case and (\textit{b}) multi-fluid case.}
		\label{fig:denpdf}
	\end{figure}
	
	From the results above, it can be stated that the variable density effects are not strongly manifested immediately after the shock wave for some of the statistics, but they play an important role in the post-shock adjustment. It is possible for these statistics, that the dominating effect across the shock is the shock compression. However, the density variations cause differences in the post-shock turbulence structure, which affects the turbulence development away from the shock wave. To get an insight into this behavior, density gradient PDFs are examined at various streamwise positions in figure \ref{fig:denpdf}. Before the shock wave, the PDFs of the density gradients are symmetric in all three directions for both single- and multi-fluid cases (not shown). For the single-fluid case, after passing through the shock wave, the density gradients' PDFs remain symmetrical, but the streamwise component PDF becomes wider due to the shock compression \citep{boukharfane2018evolution}. As the turbulence develops away from the shock wave toward the peak TKE position, the density gradients' PDFs still remain symmetrical and become narrower, which is related to the fast decay of the acoustic field. For the multi-fluid case, the density gradients' PDFs are strongly amplified through the shock wave, but the changes are relatively small far from the shock, because the density variations are controlled by the mixture composition instead of the acoustic field. More importantly, the streamwise component becomes negatively skewed. 
	
\begin{figure}
		\centering
		\includegraphics[width=5in]{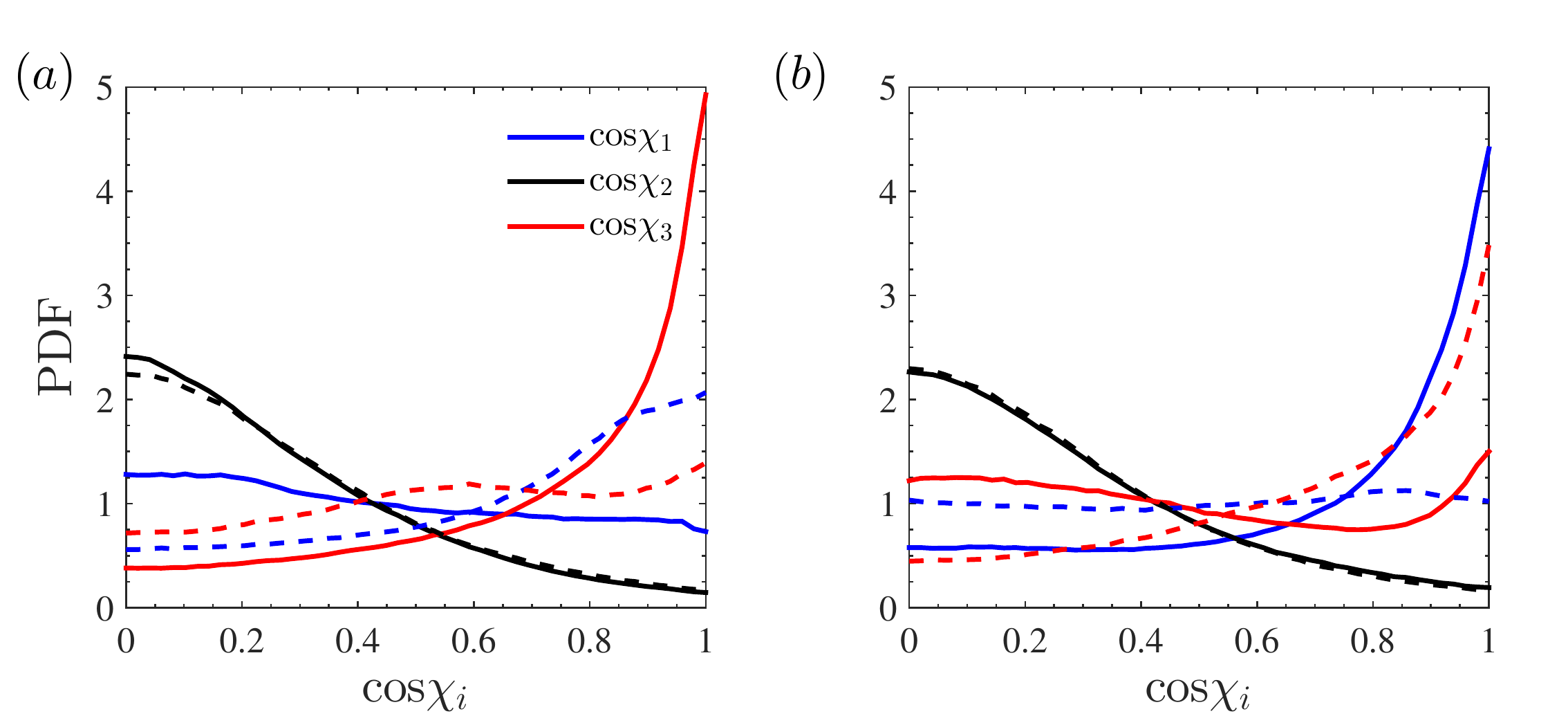}
		\caption{PDFs of the cosine angle between eigenvectors of the strain rate tensor and streamwise (x) axis for regions with (\textit{a}) $d\rho/dx > 0$ and (\textit{b}) $d\rho/dx < 0$ and for multi-fluid (solid lines) and single-fluid (dashed lines) cases.}
		\label{fig:drhoglpdf}
	\end{figure}

	To identify the mechanisms responsible for the skewness of the streamwise density gradient, we examine the orientation of the eigenvectors of strain rate tensor $S_{ij}$. The PDFs of the cosines of the angles between the three eigenvectors with the streamwise direction, conditioned on regions with positive or negative density gradients, are plotted in figure \ref{fig:drhoglpdf}. The eigenvalues of the strain rate tensor are $\gamma_1$, $\gamma_2$ and $\gamma_3$, where $\gamma_1<\gamma_2<\gamma_3$. The angles between these eigenvectors and streamwise direction are denoted by $\chi_1$, $\chi_2$ and $\chi_3$. For the multi-fluid case, in the positive density gradient regions, the extensive ($\gamma_3$-) eigenvector is more likely to be aligned with the streamwise direction (figure \ref{fig:drhoglpdf} a). The intermediate ($\gamma_2$-) eigenvector is misaligned with the streamwise direction and the compressive ($\gamma_1$-) eigenvectors have no preferential alignment. This implies that the density field is generally being stretched in the streamwise direction, making the magnitude of the density gradient smaller. On the other hand, the alignment of the $\gamma_1-$ and $\gamma_3-$ eigenvectors with the streamwise directions is reversed in the negative density gradient regions as shown in figure \ref{fig:drhoglpdf} (b). The density field is then compressed so that the magnitude of the density gradient is increased. This asymmetry in the alignment is caused by the nonlinear variable density effects when the flow passes through the shock wave and explains the negatively skewed PDF of density gradient in the multi-fluid case. It is also interesting to note the different roles of density gradient across the shock wave: spanwise density gradients contribute to the generation of the vorticity field, while the streamwise component affects the strain field. For the single-fluid case, the asymmetry in the eigenvector behavior is small and vanishes quickly away from the shock wave. This implies that even though density variations may not affect some of turbulence statistics directly, they modify the topology and structure of turbulence immediately after the shock and continue to manifest their effects in the post-shock turbulence evolution.
	
	\subsection{Topological analysis of the post-shock turbulence}
	\label{subsec:turstruc}
	
		\begin{figure}
		\centering
		\includegraphics[width=3in]{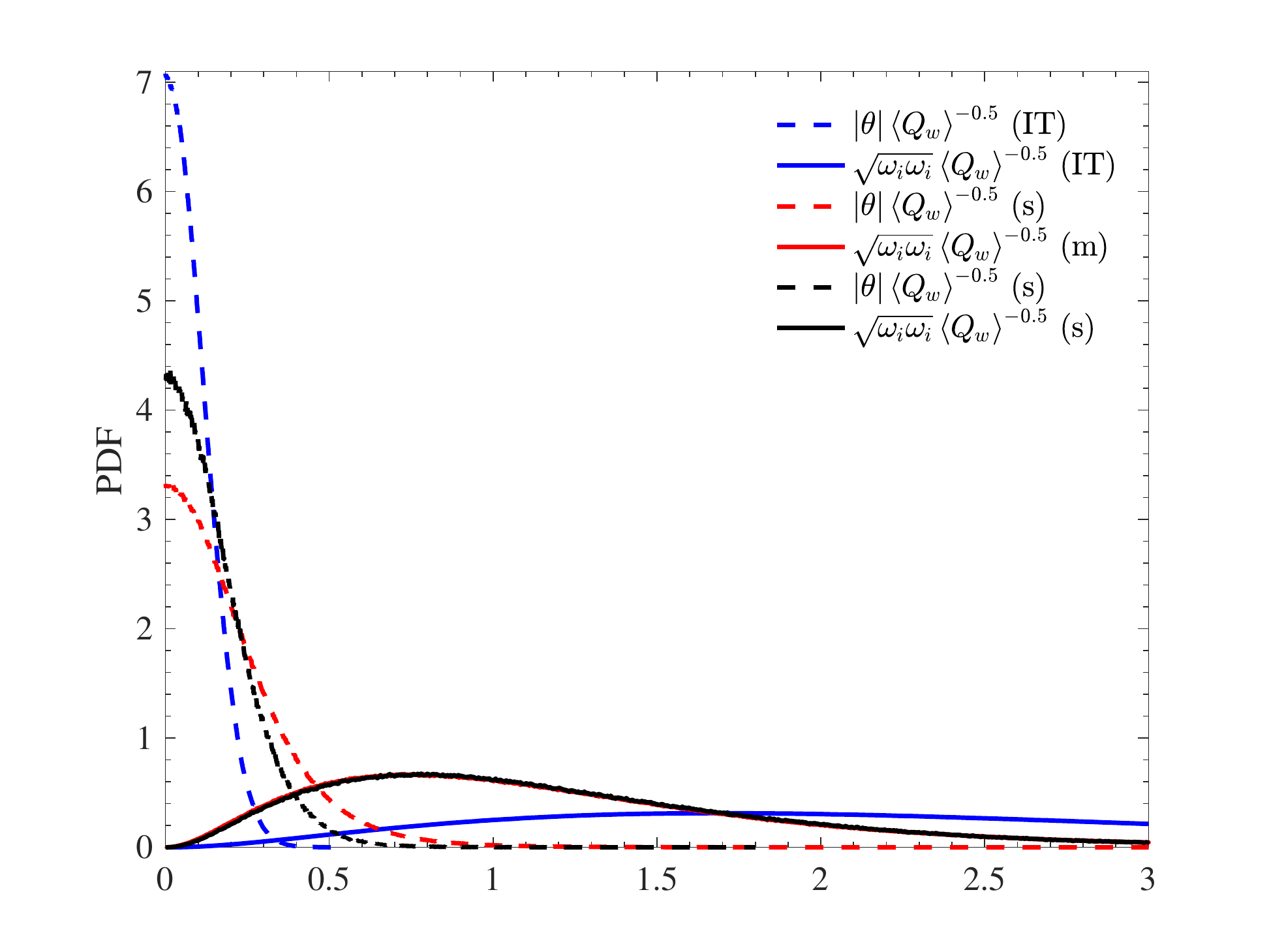}
		\caption{PDFs of the normalized dilatation and vorticity for isotropic turbulence (IT), single-fluid post-shock turbulence, and multi-fluid post-shock turbulence.}
		\label{fig:pdf_dia_omega}
	\end{figure}
	
	To further characterize the turbulence structure behind the shock wave, we have analyzed the invariant space of the VGT. The second and third invariants (denoted by $Q^\ast$ and $R^\ast$) of the anisotropic/deviatoric part of the VGT can reveal important features of the flow topology \citep{pirozzoli2004direct}. In highly compressible turbulence, there exits a richer set of flow topologies due to the dilatational part of the velocity gradient tensor \citep{suman2010velocity}. For the parameter range considered in this study; however, the compressibility effects are weak. This is demonstrated in figure \ref{fig:pdf_dia_omega}, where the normalized PDFs of the dilatation and vorticity for pre-shock isotropic turbulence, single-fluid, and multi-fluid post-shock turbulence are shown. The pre-shock isotropic turbulence has a very low magnitude of dilatation. The shock wave expectedly amplifies the dilatation magnitude, and more so when variable density effects exist, but the dilatation values are still considerably lower than those studied in \citet{suman2010velocity,chu2013topological,vaghefi2015local}. Considering that the focus of this study is on the variable density effects, here we only present the topological structure of the anisotropic velocity gradient tensor, using data points where $P\approx 0$. These regions encompass about $60\%$ of the flow. The anisotropic part of the VGT is calculated using the formula $A_{ij}^{\ast}=A_{ij} - \theta/3I$. Correspondingly, the second and third invariants can be calculated from:
	
	\begin{subeqnarray}
		\label{eqn:qr}
		Q^\ast = -\frac{1}{2}A^\ast_{ij}A^\ast_{ji}\\
		R^\ast = -\frac{1}{3}A^\ast_{ij}A^\ast_{jk}A^\ast_{ki}
	\end{subeqnarray}
	
\begin{figure}
		\centering
		\includegraphics[width=5in]{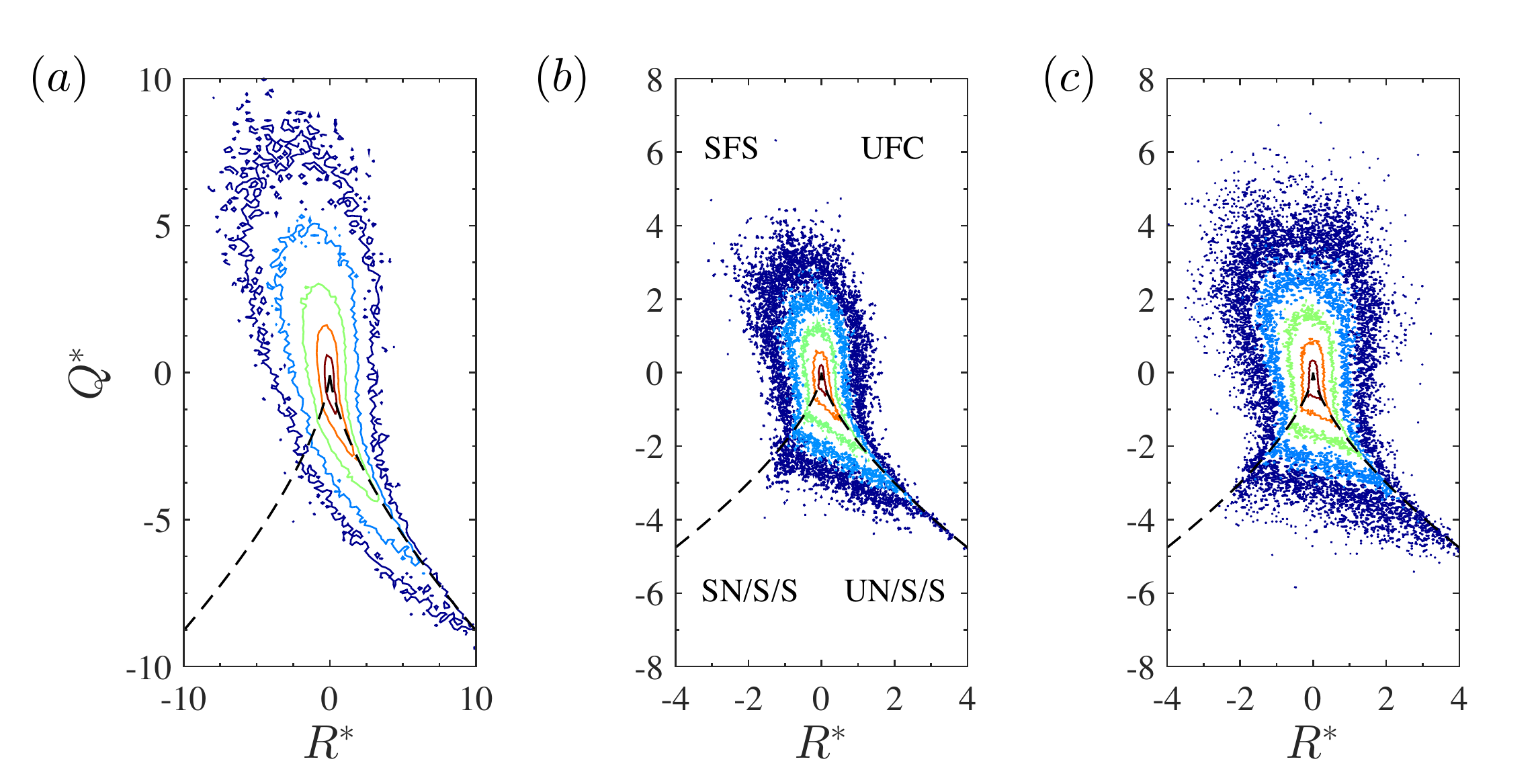}
		\caption{Iso-contour lines of joint PDFs of normalized second and third invariants of the anisotropic part of the velocity gradient tensor, ($Q^\ast$,$R^\ast$),  for (\textit{a}) pre-shock, (\textit{b}) single-fluid post-shock turbulence, and (\textit{c}) multi-fluid post-shock turbulence. The lateral lines denote the locus of zero discriminant.}
		\label{fig:QR}
	\end{figure}

	Similarly, the invariants of the symmetric and skew-symmetric parts of the anisotropic velocity gradient tensor, $S^\ast_{ij}$ and $W^\ast_{ij}$, can also be calculated using the corresponding form of equations \ref{eqn:qr}. They are denoted as ($Q_s^\ast$, $R_s^\ast$) and ($Q_w^\ast$, $R_w^\ast$) here. The following equations relate the above variables for the anisotropic part of the velocity gradient tensor \citep{ooi1999study}:
	
	\begin{subeqnarray}
		\label{eqn:qrsw}
			Q^\ast=Q_s^\ast+Q_w^\ast\\
			R^\ast=R_s^\ast-\omega^\ast_iS^\ast_{ij}\omega^\ast_j
	\end{subeqnarray}
  \noindent 
  where $\omega^\ast_i=\omega_i$ is the vorticity vector. The scalar variables $Q_s^\ast$ and $Q_w^\ast$ are related to the local dissipation rate ($-Q_s^\ast=1/2S^\ast_{ij}S^\ast_{ij}$) and enstrophy ($Q_w^\ast=1/2W^\ast_{ij}W^\ast_{ij}$), respectively. For constant viscosity, $Q^\ast$ represents the difference between enstrophy and dissipation \citep{chu2013topological}. Similarly, $R_s^\ast=-1/3S^\ast_{ij}S^\ast_{jk}S^\ast_{ki}$ is related to the production of dissipation due to strain field and $\omega^\ast_iS^\ast_{ij}\omega^\ast_j$ is the vortex stretching contribution to the enstrophy. Therefore, for constant viscosity, $R^\ast$ represents the difference between enstrophy production and dissipation production. Based on the local values of $Q^\ast$ and $R^\ast$, four types of local flow topologies can be identified: stable-focus/stretching (SFS), unstable-focus/contracting (UFC), stable-node/saddle/saddle (SN/S/S) and unstable-node/saddle/saddle (UN/S/S). For isotropic turbulence, the joint PDF of ($Q^\ast,R^\ast$) has the tear-drop shape. This has been further observed in other fully developed turbulent flows, such as boundary layers, mixing layers, and channel flows \citep{pirozzoli2004direct,wang2012effect}. This type of distribution of $Q^\ast$ and $R^\ast$ is an indicator that the turbulence is more likely having a local topology of stable-focus/stretching or an unstable-node/saddle/saddle. In figure \ref{fig:QR} (a), it is shown that the joint PDF of normalized second and third invariants, $Q^\ast/ \langle Q_w \rangle$ and $R^\ast/\langle Q_w \rangle^{3/2}$, has the same tear-drop shape in the pre-shock flow. Using shock-LIA and DNS data, \citet{ryu2014} showed that for single-fluid STI, the ($Q^\ast,R^\ast$) distribution is significantly modified by the shock wave, with a tendency towards symmetrization of the joint PDF. This indicates that the regions with stable-focus/compression and stable-node/saddle/saddle (first and third quadrant) are more likely to occur as turbulence develops a 2-D axisymmetric flow structure. To understand the variable density effects on this shock-induced symmetrization, the joint PDFs of ($Q^\ast$,$R^\ast$) for both single-fluid and multi-fluid post-shock turbulence are compared in figure \ref{fig:QR} (b,c).
	
	Figure \ref{fig:QR} (b) shows the joint distribution for the single-fluid post-shock turbulence. The dashed lines denote the locus of zero discriminant of $A^{\ast}$, where $Q^\ast$ and $R^\ast$ satisfy $27R^{\ast 2}/4+Q^{\ast 3}=0$. Compared to the pre-shock joint PDF, there is a tendency towards symmetrization, with more points located in the first and third quadrants. Similar to single-fluid STI, multi-fluid STI demonstrates a tendency towards symmetrization of the $(Q^\ast, R^\ast)$ distribution. However, the multi-fluid distribution is slightly more symmetric and has a larger variance, with more points away from the axes. This implies that more extreme "events" exist in the post-shock multi-fluid turbulence.

\begin{figure}
		\centering
		\includegraphics[width=5in]{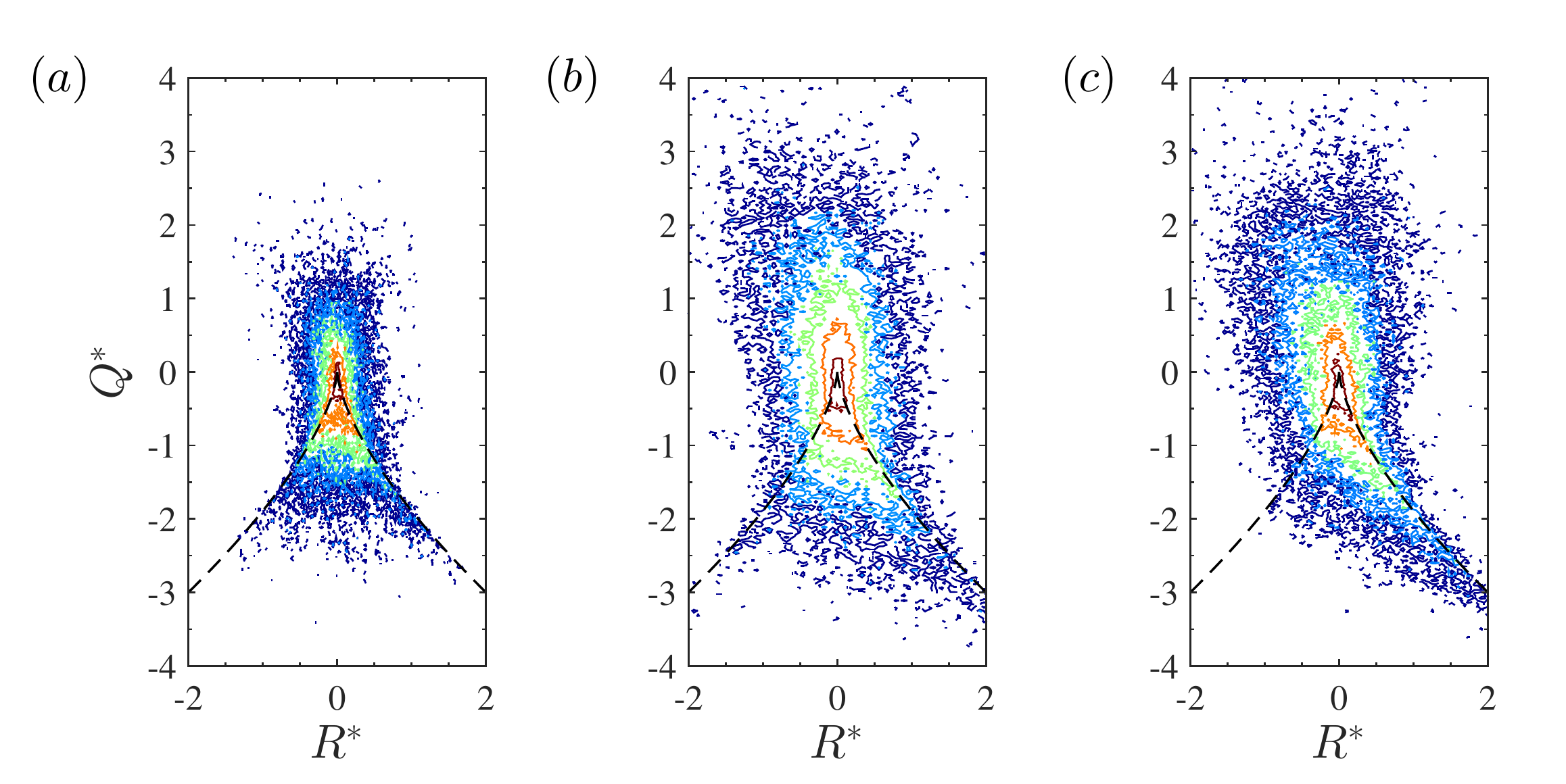}
		\caption{Iso-contour lines of post-shock ($k_{0}x \approx 0.44$) joint PDF of second and third invariants of the anisotropic part of the velocity gradient tensor, ($Q^\ast$, $R^\ast$),  in regions with different densities. (\textit{a})  regions with high density values, $\rho>(\overline{\rho}+90\%\rho'_{rms})$,  (\textit{b}) regions with density around the post-shock mean value, and (\textit{c})  regions with low density values, $\rho<(\overline{\rho}-90\%\rho'_{rms})$. }
		\label{fig:QRden}
	\end{figure}

	The density effects on the post-shock joint PDF of second and third invariants are further explored by comparing the conditional distribution, conditioned on regions with different densities, in figure \ref{fig:QRden} (a)-(c). Figure \ref{fig:QRden} (a) corresponds to regions with relatively high density ($\rho>(\overline{\rho}+90\%\rho'_{rms})$), \ref{fig:QRden} (b) to regions with density around the post-shock mean value, and \ref{fig:QRden} (c) to low density regions ($\rho<(\overline{\rho}-90\%\rho'_{rms})$). For consistency check, the joint PDFs corresponding to these regions are also computed for the pre-shock flow (not shown) and found to be close to the single-fluid PDFs. After the shock wave, the joint PDFs demonstrate significant differences between regions with different densities. In regions with density closer to that of the post-shock mean density, the distribution of invariants appears to be very similar to that shown in figure \ref{fig:QR} (c). But for regions with higher density (figure \ref{fig:QRden} (a)), the joint PDF becomes more symmetric compared to the overall flow or single-fluid case. There is a much larger portion of data points having a local topology of stable-node/saddle/saddle, and fewer data points fall into the first and second quadrants, indicating larger strain-dominated regions. On the other hand, the post-shock regions with low-density values (figure \ref{fig:QRden} (c)) exhibit features similar to that of isotropic turbulence, with almost the same tear-drop shape, only with a larger variance or a wider distribution. The quantitative difference is hypothesized to be related to the higher shock strength variation in the multi-fluid case. It was observed in our previous studies \citep{tian2019shock}, that the local shock strength is positively correlated with the pre-shock density. With a stronger shock, the two-dimensionalization effect on the post-shock turbulence should also appear stronger in the high-density regions
	\citep{livescu2015vorticity}. For low-density regions, the smaller two-dimensionalization effect reduces the symmetrization trend. Moreover, the relatively lower inertia in these regions leads to a faster response to the local strain field \citep{livescu2010new}, which could make a faster return to isotropic turbulence. The different characteristics of ($Q^\ast, R^\ast$) joint PDF in regions with different densities provide additional evidence for the previous argument made about the density role on the preferential amplification of the strain and rotation tensors.

	\begin{figure}
		\centering
		\includegraphics[width=5in]{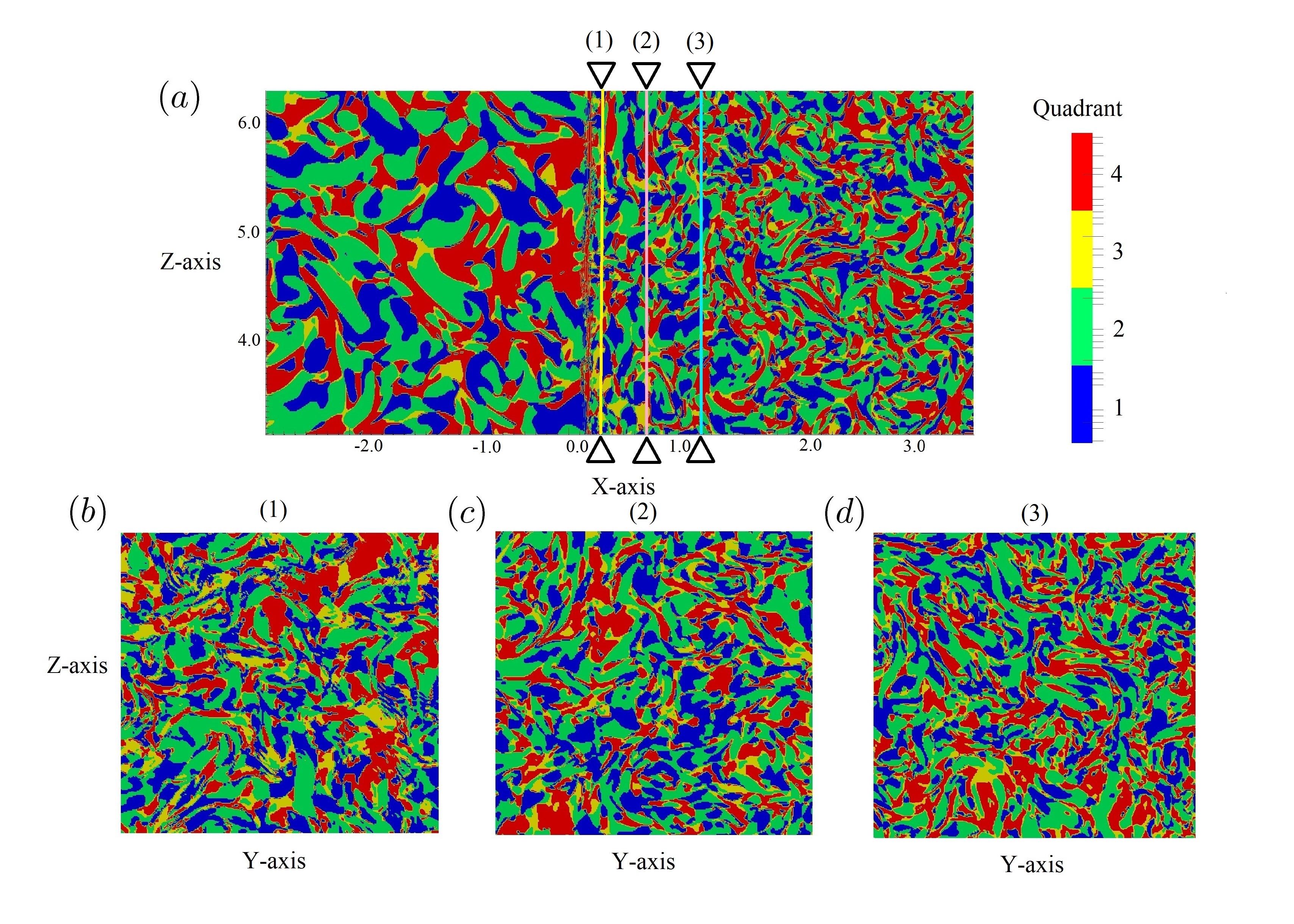}
		\caption{Color illustration of the flow topology for the multi-fluid STI. The flow topology is represented by the quadrants (denoted as $Q_i$) of the joint PDF of ($Q^\ast$, $R^\ast$). (\textit{a}) 2D color contours in the x-z plane at $y=3.14$ (half y-domain). The shock wave is located in the middle of the domain at $k_0x \approx 0.0$. The ratio of the fluid volume in different quadrants in the pre-shock region is $Q_1:Q_2:Q_3:Q_4$=26.7\%:38.7\%:7.8\%:26.8\%. The 2D color contours in the homogeneous y-z plane at streamwise locations of: (\textit{b}) $k_0x \approx 0.2$ (28.7\%:34.4\%:14.3\%:22.6\%), (\textit{c}) $k_0x \approx 2.0$, peak TKE location (26.7\%:36.9\%:11.2\%:25.2\%) (\textit{d}) $k_0x \approx 4.0$ (26.3\%:37.9\%:9.3\%:26.2\%).}
		\label{fig:quadrant}
	\end{figure}
	
	In figure \ref{fig:quadrant}, the planar distribution of the flow topologies are shown. Here, $Q_i$ refers to the quadrants on the joint PDF of ($Q^\ast$, $R^\ast$), which amounts to a representation of the local flow topology. Figure \ref{fig:quadrant} (a) presents the 2D visualization of the flow topology in a typical $x-z$ plane. The regions occupied by different quadrants are marked using different colors. Evidently, the vorticity-dominated regions ($Q_1$ and $Q_2$) cover a large portion of the flow and have more compact shapes. These regions are connected by UN/S/S areas ($Q_4$), which are more elongated. The SN/S/S ($Q_3$) areas can be located either at the edge of $Q_4$ regions or in-between different $Q_4$ regions. These strain-dominated regions seem to be more fragmented than the compact vorticity-dominated regions. Figure \ref{fig:quadrant} (b)-(d) show the 2D contours of $Q_i$ in the homogeneous ($y-z$) planes at different streamwise locations after the shock. These locations are also marked on figure \ref{fig:quadrant} (a). Immediately after passing through the shock wave, the volume fractions of different quadrants are calculated to be $Q_1:Q_2:Q_3:Q_4=28.7\%:34.4\%:14.3\%:22.6\%$, indicating a trend towards symmetrization in the joint PDF. This can also be observed in the 2D contours in figure \ref{fig:quadrant} (b). Moreover, the characteristic length scales associated with the regions occupied by different quadrants are decreased across the shock wave. As the flow evolves away from the shock wave, the distribution slowly changes back to the pre-shock shape but still with smaller turbulence length scales. Most of fluid in different quadrants return to the pre-shock values at $k_0x=4.0$. The re-orientation of the flow structures into the streamwise direction is also noted in figure \ref{fig:quadrant} (a), consistent with the return to isotropy of the flow. However, the rates at which different flow features return to an isotropic state are slightly different. The dynamics of flow and the return-to-isotropic turbulence process are examined in detail in the next section using the Lagrangian statistics.
	
 The quasi-axisymmetric state immediately after the shock wave, identified above based on the joint PDF of ($Q^\ast$, $R^\ast$), is further explored below by considering the vortex stretching rate and other flow topological features.
	
\begin{figure}
		\centering
		\includegraphics[width=5in]{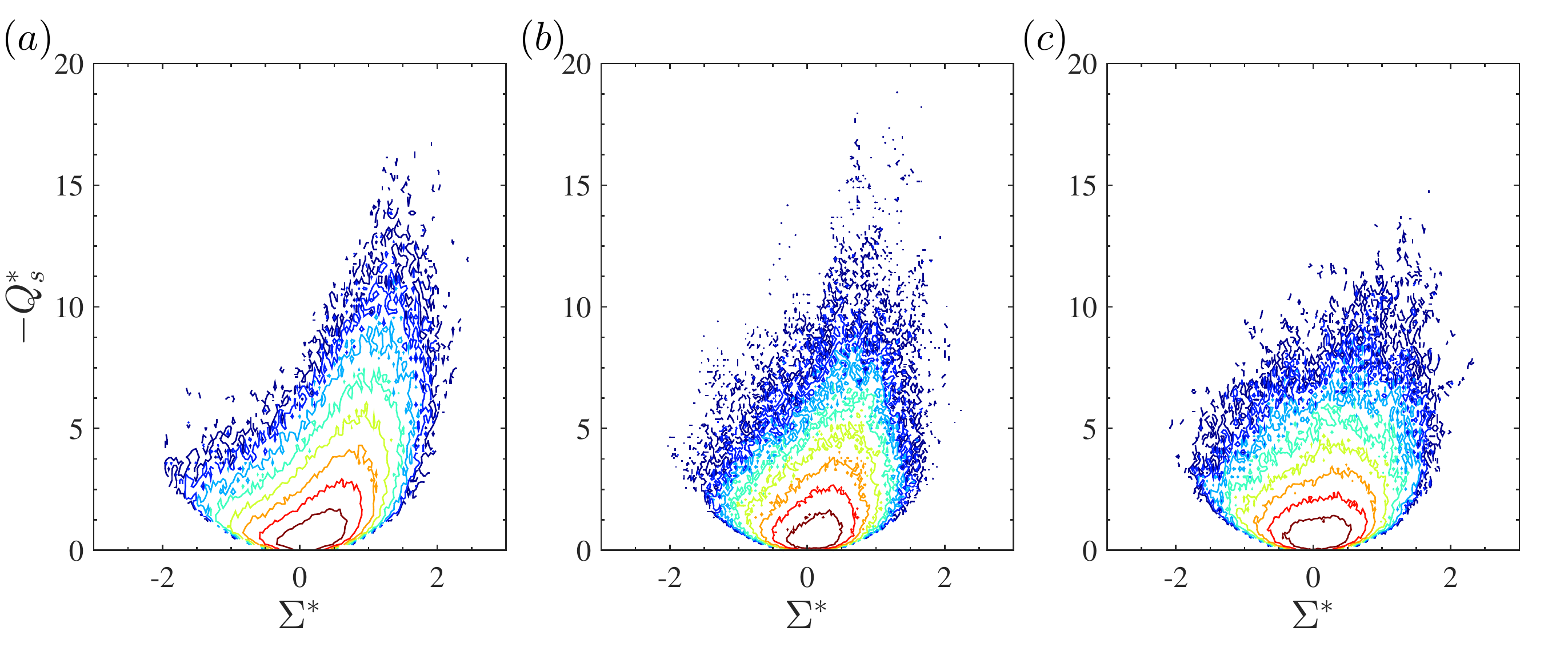}
		\caption{Iso-contour lines of joint PDF of (-$Q_s^\ast$, $\Sigma^\ast$) for (\textit{a}) isotropic box turbulence and (\textit{b,c}) single-fluid and multi-fluid turbulence at post-shock position of $k_{0}x \approx 0.44$.}
		\label{fig:SigmaQs}
	\end{figure}

	The rate at which the vorticity is stretched or contracted, i.e. the normalized vortex stretching rate, can be calculated based on the VGT invariants using the formula: $\Sigma^\ast =w_iS_{ij}w_j= (R_s^\ast-R^\ast)/Q_w^\ast$ \citep{ooi1999study}. In figure \ref{fig:SigmaQs}, the joint PDF of ($-Q_s^\ast$, $\Sigma^\ast$) is plotted to investigate the effects of the strain field on the vortex stretching rate. Positive and negative $\Sigma^\ast$ values correspond to the vortex being stretched or contracted. Figure \ref{fig:SigmaQs} (a) shows the joint PDF of ($-Q_s^\ast$, $\Sigma^\ast$) for the isotropic turbulence. The results agree very well with those of \citet{ooi1999study}, which indicates that the flow favors positive $\Sigma^\ast$ values or an overall vortex stretching, especially in the strong strain dominated regions. Here, we compare the results from isotropic turbulence to those from single-fluid and multi-fluid post-shock turbulence to understand the shock and variable density effects. We note that in figure \ref{fig:SigmaQs} (b), the joint PDF becomes more symmetric around $\Sigma^\ast=0.0$ after passing through the shock wave.  For the multi-fluid case, as shown in figure \ref{fig:SigmaQs} (c), the joint PDF becomes almost fully symmetric, especially at lower $-Q_s^\ast$ values. This symmetry has a strong effect on the overall vortex stretching rate for the multi-fluid post-shock turbulence because the positive and negative $\Sigma^\ast$ values tend to cancel each other through averaging. Moreover, the variances of the stretching term are almost the same for single and multi-fluid cases, meaning that the lower stretching rate is mainly due to changes in the turbulence structure (especially in more negative $\Sigma^\ast$ regions), and not simply the decrease in the magnitude of $\Sigma^\ast$. 
	
\begin{figure}
		\centering
		\includegraphics[width=5in]{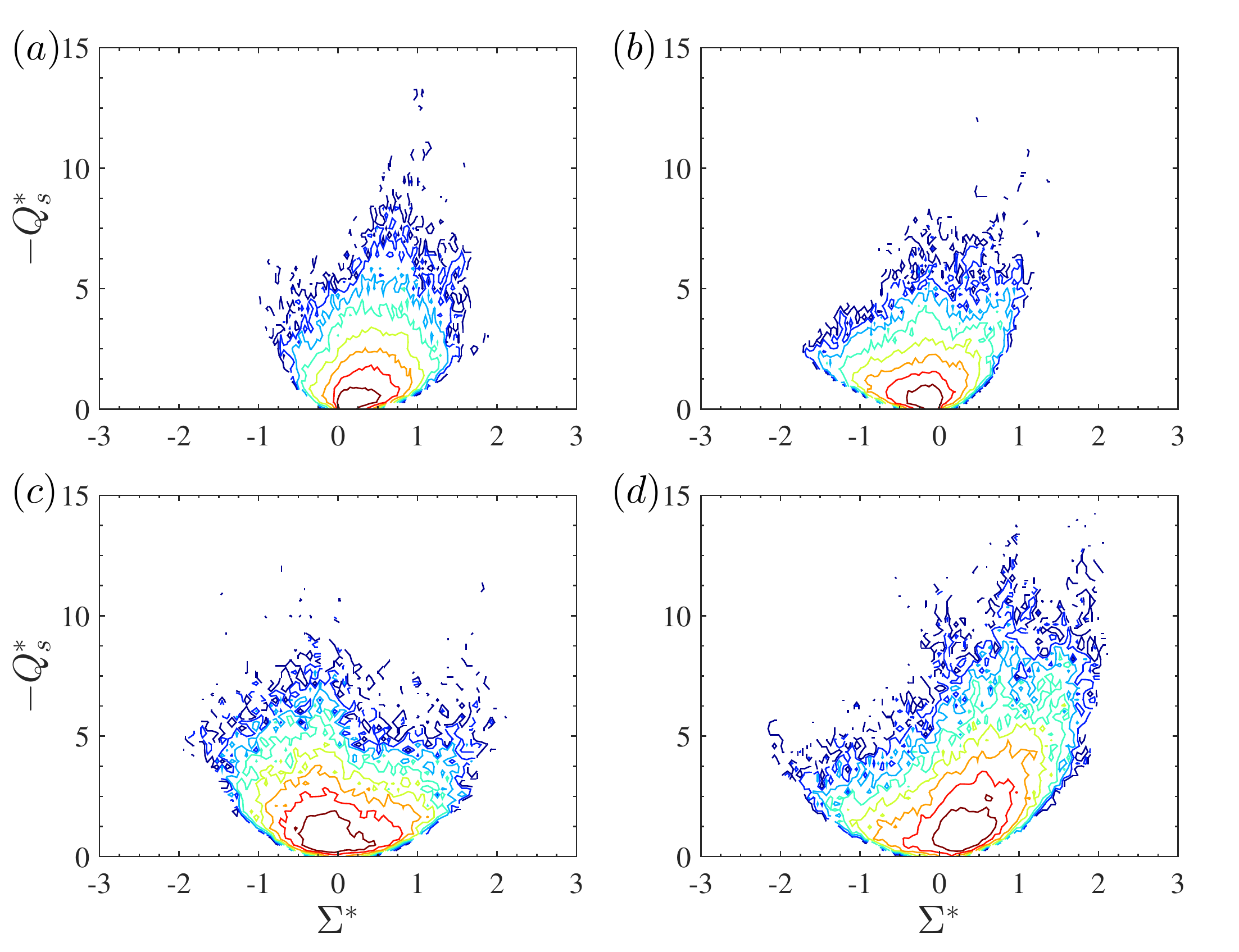}
		\caption{Iso-contour lines of joint PDF of (-$Q_s^\ast$, $\Sigma^\ast$) for different quadrants right after the shock wave. (\textit{a}) $Q_{2}$, (\textit{b}) $Q_{1}$ ,(\textit{c}) $Q_{3}$ and ($\textit{d}$) $Q_{4}$.}
		\label{fig:SigmaQsquadrant}
	\end{figure}

	To understand the contribution from different topological states to the vortex stretching, the joint PDF of ($-Q_s^\ast$, $\Sigma^\ast$) is conditioned on different quadrants for the multi-fluid case. Figure \ref{fig:SigmaQsquadrant} (a,b) shows the joint PDF for $Q_1$ and $Q_2$ regions, or areas with a local topology of SFS and UFC. It can be observed that in these rotation-dominated regions, the magnitude of the vortex stretching rate $\Sigma^\ast$ is relatively small. Moreover, $Q_1$ is dominated by areas with negative $\Sigma^\ast$ and $Q_2$ is dominated by positive $\Sigma^\ast$ areas, which can be inferred from their corresponding topologies. However, for $Q_3$ and $Q_4$ (figure \ref{fig:SigmaQsquadrant} c,d), the magnitude of the vortex stretching rate is larger than that in the rotation-dominated regions (figure \ref{fig:SigmaQsquadrant} a,b). In $Q_3$, the joint PDF is relatively symmetric and seems to be slightly biased towards negative vortex stretching rate values. $Q_4$, on the other hand, is dominated by positive vortex stretching. Overall, the results explain the lower averaged vortex stretching rate values in the multi-fluid case caused by the enhancement of $Q_1$ and $Q_3$ regions.

	\subsection{Lagrangian dynamics of VGT}
	\label{subsec:lagrangian}
	In this section, we use non-inertial Lagrangian particles/tracers that move with the local flow velocity because their statistics can reflect important transient turbulent dynamics, which are difficult to study using Eulerian data \citep{yeung2002lagrangian}. More importantly, in the context of variable density flows, the Lagrangian statistics enable us to differentiate among particles with different densities and investigate their dynamics separately. Lagrangian data are used here to perform an analysis of the post-shock turbulence structure and VGT dynamics with and without significant density fluctuations. 
	
\begin{figure}
		\centering
		\includegraphics[width=5in]{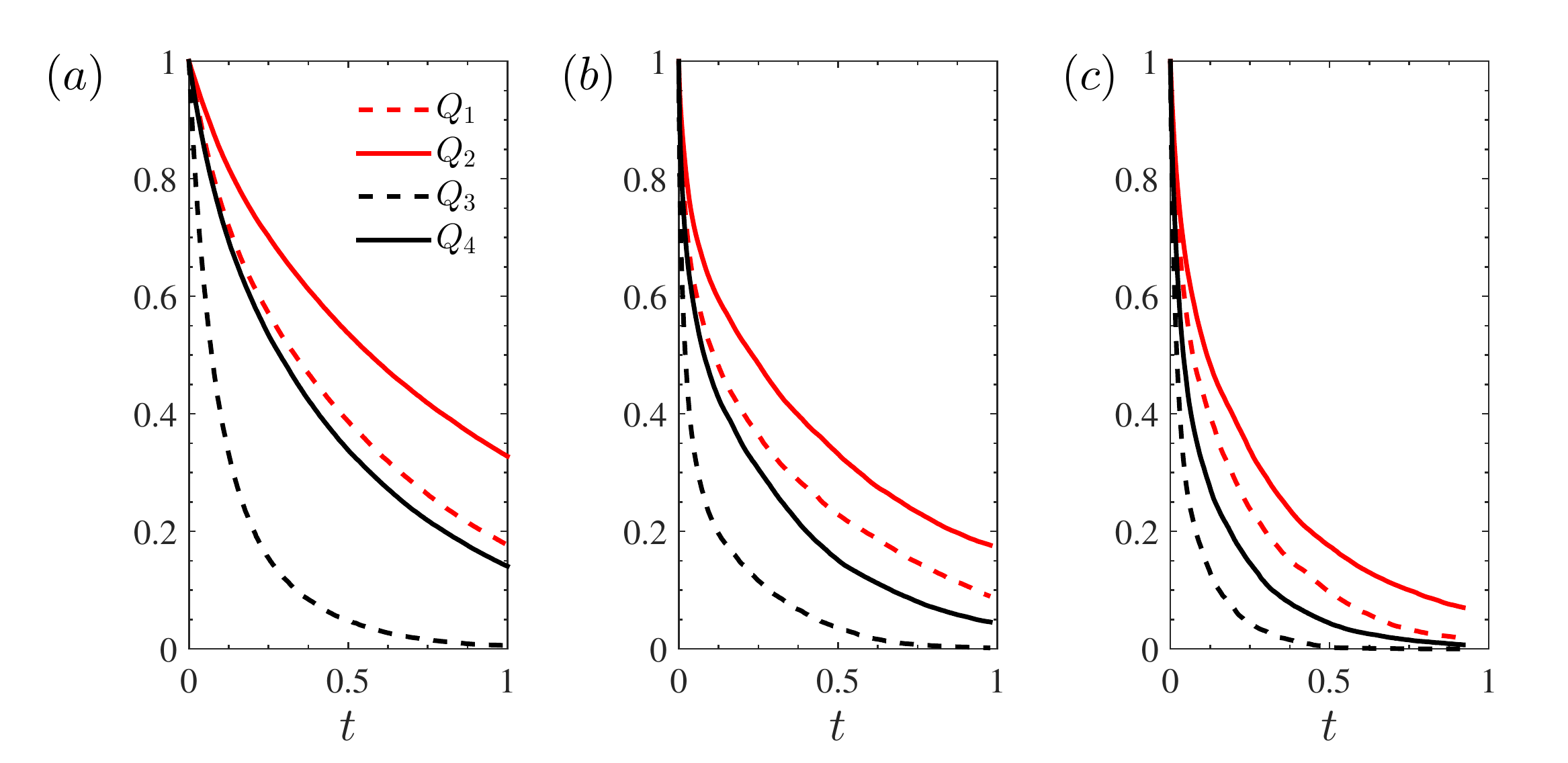}
		\caption{Percentage of fluid particles that stay in each quadrant following  particles initialized uniformly in (\textit{a}) isotropic turbulence and (\textit{b,c}) single-fluid and multi-fluid turbulence at post-shock position of $k_{0}x=0.44$.}
		\label{fig:figure13}
	\end{figure}

	The first result considered is the timescale of particle motions related to different flow topologies. In figure \ref{fig:figure13}, the percentage of fluid particles that remain in their starting quadrants are plotted over time so that we can identify the residence time of particles for different turbulence structures. In figure \ref{fig:figure13} (a), the percentages of fluid particles are plotted for decaying isotropic turbulence as a reference. It is noted that $Q_{3}$ and $Q_{4}$, which are the strain-dominated regions, have the smallest associated residence times among all the four quadrants, with $Q_{3}$ time being the smaller of the two. For the rotation-dominated regions, the residence times are expectedly longer, especially for $Q_{2}$. The residence times for single-fluid and multi-fluid simulations can be inferred from figure \ref{fig:figure13} (b,c). For both cases, $Q_{3}$ always has the least residence time and $Q_{2}$ has the largest one. Comparing all three cases, the particles in the multi-fluid and single-fluid post-shock turbulence are shown to evolve faster away from the original quadrant than particles in isotropic turbulence, indicating smaller timescales of the flow topologies. Between the two post-shock turbulence fields, the multi-fluid case presents shorter residence times.
	
\begin{figure}
		\centering
		\includegraphics[width=5in]{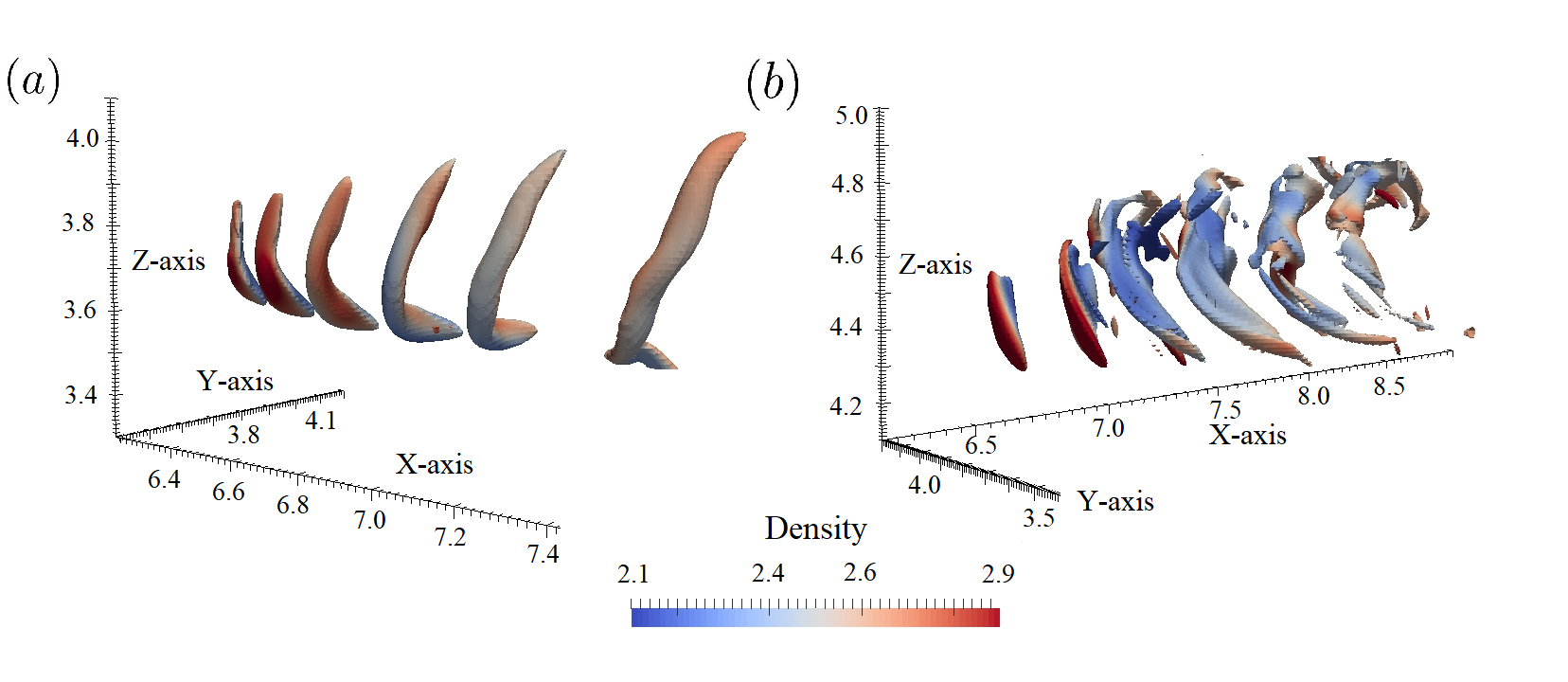}
		\caption{Visualization of the temporal development (left to right) of the turbulence structure using iso-surfaces of $Q^\ast$ colored by density for the multi-fluid post-shock turbulence. These structures are captured immediately after the shock wave. (\textit{a}) vorticity-dominated structure, and (\textit{b}) strain-dominated structure. }
		\label{fig:vortexvis}
	\end{figure}

	 Figure \ref{fig:vortexvis} presents an example of the temporal development of the above-mentioned structures. The evolution of a vortex tube in the post-shock turbulence is tracked and visualized as it moves away from the shock wave in figure \ref{fig:vortexvis} (a). As expected, the depicted vortical structure maintains its shape, except that it is being stretched and reoriented by the local flow field. Moreover, the vortex tube surface is almost parallel with the iso-surface of the density field, i.e. perpendicular to the density gradient. This is consistent with the discussion in section \ref{subsubsec:turbstate} regarding the bulk of vorticity generation across the shock wave. As the vortex tube evolves away from the shock wave, the reorientation of the density gradient by the vortex is also observed. In figure \ref{fig:vortexvis} (b), a strain-dominated structure is visualized using the iso-surface of negative $Q^\ast$. It can be clearly seen that such structures lack temporal coherency since they tend to be become fragmented as they evolve.

	\begin{figure}
		\centering
		\includegraphics[width=5in]{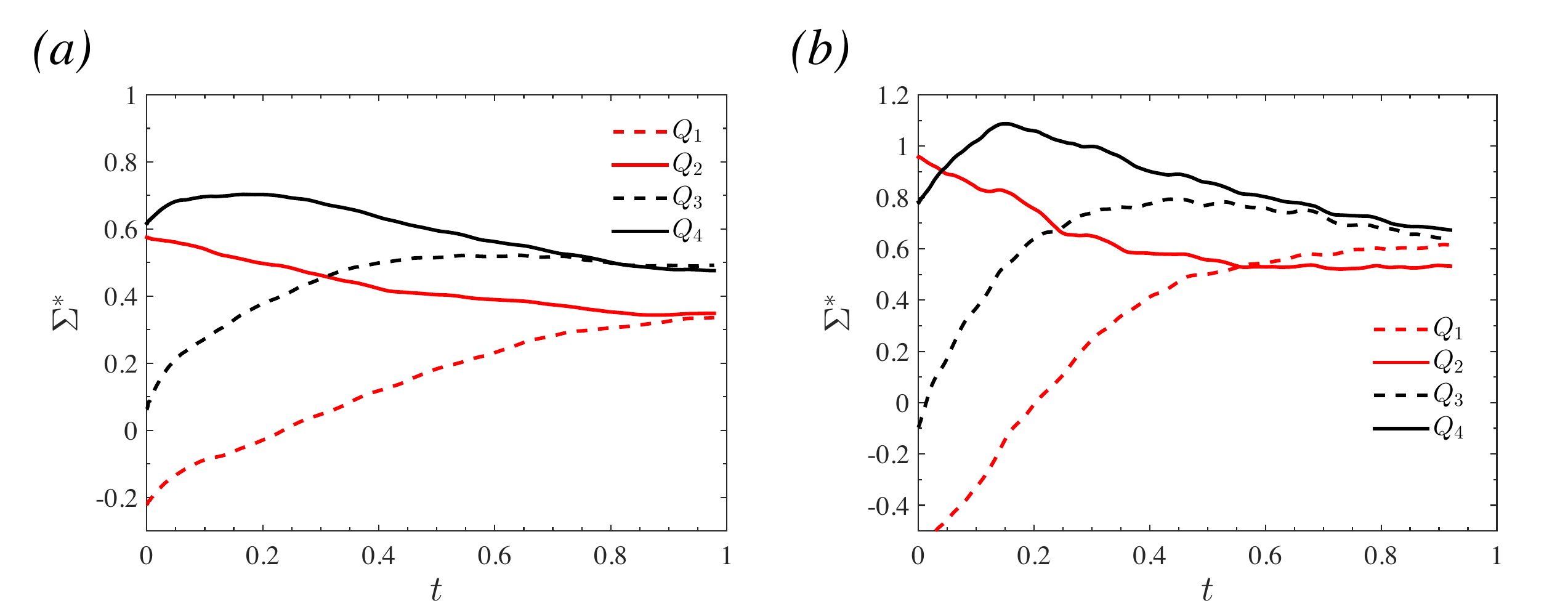}
		\caption{Contributions to the vortex stretching rate from particles starting in each quadrant. The particles are initialized uniformly at the post-shock position $k_{0}x \approx 0.44$ and traced downstream till the vorticity returns to an isotropic state. (\textit{a}) single-fluid and (\textit{b}) multi-fluid cases.}
		\label{fig:figure14}
	\end{figure}
	
	In figure \ref{fig:figure14}, the contributions to the normalized vortex stretching rate from particles that are initialized in each of the four quadrants are plotted following these particles. As expected, at $t=0.0$, particles from $Q_2$ and $Q_4$ add positively to the vortex stretching rate, while those from $Q_1$ have a negative vortex stretching rate contribution on average. This is in good agreement with the joint PDFs of ($-Q_s^\ast$, $\Sigma^\ast$), shown in figure \ref{fig:SigmaQsquadrant}. For $Q_3$, the initial contribution is close to zero. As the fluid particles move with the turbulent flow, their contributions to the vortex stretching also change. It can be seen that the fast increase in vortex stretching can be mainly attributed to the particles originating in $Q_1$ and $Q_3$. Particles starting in $Q_4$ have an increasing vortex stretching contribution for a short period before their combined/average contributed value starts to decrease. The behavior is qualitatively similar for the single-fluid case, but the changes are smaller in this case. For both cases, the vortex stretching contribution from the initial $Q_2$ particles decreases in time.

	To further understand this behavior, the Lagrangian equations of the VGT and its invariants are considered. The time evolution of $A_{ij}$ for fluid particles can be obtained by taking the spatial derivatives of the Navier-Stokes equations. In dimensionless form, it can be written as \citep{chu2013topological}:
	
	\begin{subeqnarray}
		\frac{\partial A_{ij}}{\partial t} + u_{k} \frac{\partial A_{ij}}{\partial x_{k}} + A_{ik}A_{kj}=-H_{ij}+\mathcal{T}_{ij}\\
		\frac{D A_{ij}}{D t} =- A_{ik}A_{kj} -H_{ij}+\mathcal{T}_{ij}
	\end{subeqnarray}
	
	with
	
	\begin{subeqnarray}
		H_{ij}&=&\frac{\partial}{\partial x_{j}} (\frac{1}{\rho}\frac{\partial p}{\partial x_{i}})=-\frac{1}{\rho^2}\frac{\partial \rho}{\partial x_{j}}\frac{\partial p}{\partial x_{i}} + \frac{1}{\rho}\frac{\partial p^2}{\partial x_{i}\partial x_{j}}=H^b_{ij}+H^p_{ij} \\
		\mathcal{T}_{ij}&=&\frac{\partial}{\partial x_{j}}(\frac{1}{\rho}\frac{\partial \sigma_{ik}}{\partial x_k})\\
		\sigma_{ij}&=&\frac{\mu}{\Rey_{0}} \left ( \frac{\partial u_{i}}{\partial x_{j}}+ \frac{\partial u_{j}}{\partial x_{i}} -\frac{2}{3} \frac{\partial u_{k}}{\partial x_{k}} \delta_{ij} \right )
	\end{subeqnarray}
	
	where $Re_0$ is the reference Reynolds number. From here, the dynamic equations for the three invariants of the VGT, $P$, $Q$, and $R$ can be derived in the following form \citep{chu2013topological}:
	
	\begin{subeqnarray}
		\frac{D P}{D t}&=&(P^{2}-2Q) + H^p_{ii} + H^b_{ii}- \mathcal{T}_{ii}\\
		\frac{D Q}{D t}&=&(PQ-3R) + (PH^p_{ii}+A_{ij}H^p_{ji})+ (PH^b_{ii} +A_{ij}H^b_{ji})+(- P\mathcal{T}_{ii}-A_{ij}\mathcal{T}_{ji})\\
		\frac{D R}{D t}&=&PR + (QH^p_{ii}+PA_{ij}H^p_{ji}+A_{ij}A_{jk}H^p_{ki})+ (QH^b_{ii}+PA_{ij}H^b_{ji} \nonumber \\
		&&+A_{ij}A_{jk}H^b_{ki}) +(- Q\mathcal{T}_{ii}-PA_{ij}\mathcal{T}_{ji}-A_{ij}A_{jk}\mathcal{T}_{ki})
		\label{eqn:lagdynamic}
	\end{subeqnarray}
	
	where the three invariants of VGT are defined as:
	
		\begin{subeqnarray}
		\label{eqn:qrvgt}
			P &=& -tr(A_{ij})\\	
		Q &=& \frac{1}{2}(tr(A_{ij})^2-tr(A_{ij}A_{jk}))\\
		R &=& -det(A_{ij})
	\end{subeqnarray}

	Here, $tr(A_{ij})$ and $det(A_{ij})$ denote the trace and determinant of a tensor. Note that instead of the deviatoric part of the VGT, the dynamic equations for the full VGT are considered. The reason is that due to the variable density effects and shock compression, the incompressibility condition is not satisfied especially when $M_t$ and $A_t$ become large. Even though $M_t$ and $A_t$ in this study are small, we still consider the full equations for any future comparisons. The contributions from the dilatational part of the VGT and their coupling with the variable density effects in highly compressible turbulence are still unknown and need to be explored for STI in future studies.
	
	The dynamical equations can be divided into contributions by four different parts: I) mutual-interaction among invariants, II) pressure Hessian, $H^p_{ij}$, III) baroclinic, $H^b_{ij}$ and IV) viscous term $\mathcal{T}_{ij}$. The statistics regarding these terms can be extracted from the Lagrangian data.

	\begin{figure}
		\centering
		\includegraphics[width=5in]{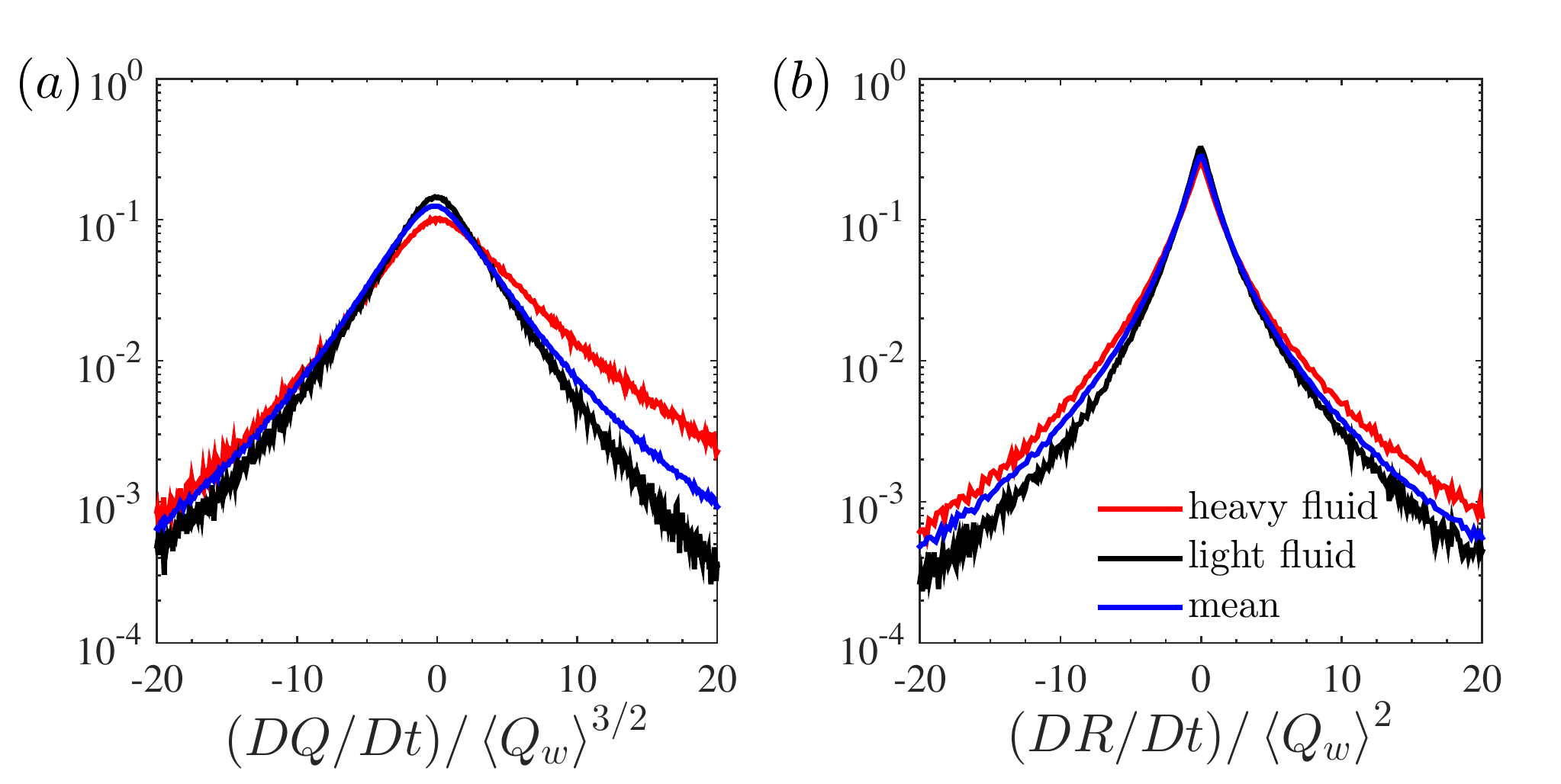}
		\caption{PDFs of (\textit{a}) $(DQ/Dt)/\left \langle Q_w \right \rangle ^{3/2}$ and (\textit{b}) $(DR/Dt)/\left \langle Q_w \right \rangle ^{2}$ for fluid particles with different densities at streamwise location of $k_0x \approx 0.5$.}
		\label{fig:dqdrmhl}
	\end{figure}

	Some general features of the Lagrangian dynamics of the VGT invariants are examined through the PDFs of their material derivatives. The variable density effects can be identified by comparing the PDFs corresponding to regions with different densities (figure \ref{fig:dqdrmhl}). In the light fluid regions, the PDFs of $DQ/Dt$ and $DR/Dt$ have narrower tails, while the tails are wider in the heavy fluid regions. Another important observation is that the skewness of $DQ/Dt$ is different in the light and heavy fluid regions. Heavy fluid particles have a positively-skewed PDF, similar to the overall flow. On the other hand, the $DQ/Dt$ skewness resulting from light fluid particles is negative. This implies that heavy fluid particles are more likely to move towards rotation-dominated regions and vice versa. These differences can be attributed to differences in the return-to-isotropy, experienced by fluid particles with different densities. 
	
	\begin{figure}
		\centering
		\includegraphics[width=5in]{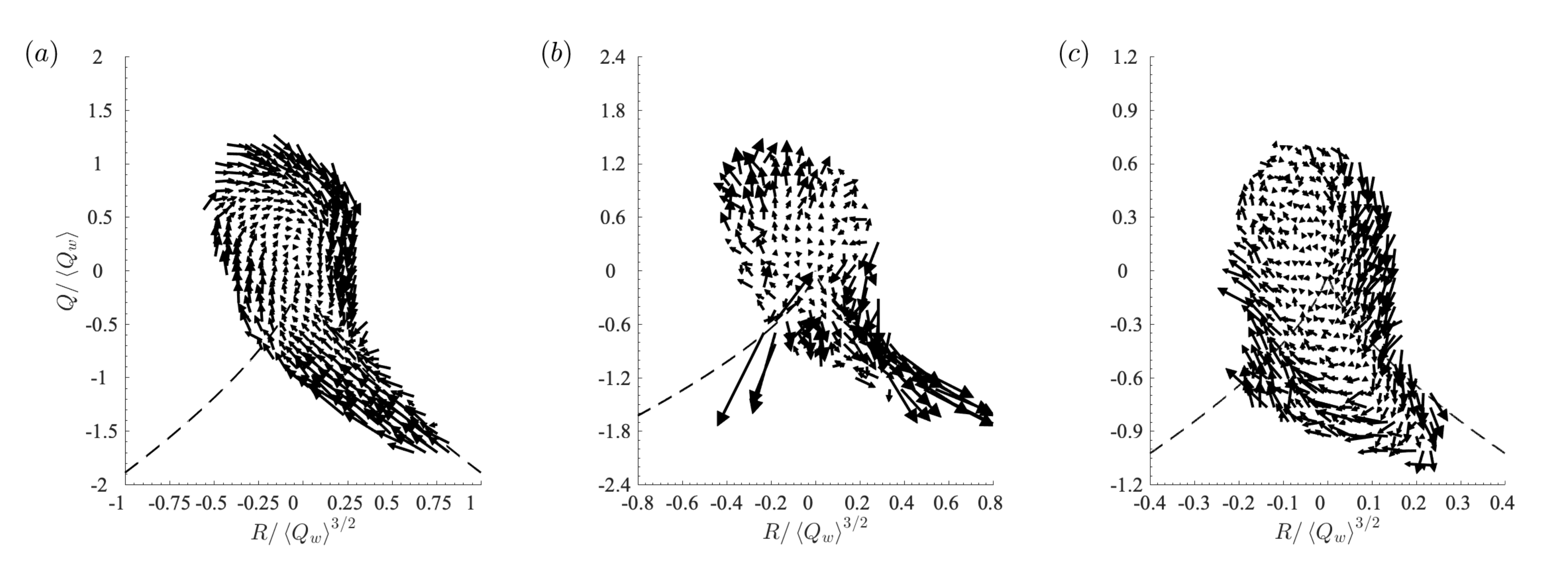}
		\caption{Conditional mean rate of change vectors of ($DQ/Dt/\left \langle Q_w \right \rangle ^{3/2}$,$DR/Dt/\left \langle Q_w \right \rangle ^{2}$)  in the ($Q, R$) plane for (\textit{a}) isotropic turbulence, (\textit{b}) single-fluid post-shock turbulence, and (\textit{c}) multi-fluid post-shock turbulence at streamwise location of $k_0x \approx 0.5$. To ensure that the vectors can be properly visualized, their sizes are re-scaled by multiplying with a constant of 0.3. This applies to all the following vector plots.}
		\label{fig:vectorQR}
	\end{figure}

	The Lagrangian dynamics of the turbulence and the evolution of flow topology are further examined here by considering the conditional mean rate of change of $Q$ and $R$ in the invariants plane \citep{ooi1999study}. The rates of change are used to form a vector at each point in the invariants plane. The trajectories implied by these vectors can be followed to understand the return-to-isotropy process. In fully compressible turbulence, the $(P, Q, R)$ invariant space becomes three-dimensional \citep{suman2010velocity, chu2013topological,vaghefi2015local} and there exists an out-of-plane $(Q, R)$ component of the trajectory due to the contribution from compressibility ($P$) effect. Due to the low compressibility effect in this work, however, it would be more appropriate to consider only the in-plane $(Q, R)$ dynamics and leave the compressibility effects for future study. Therefore, the results presented below correspond to the data points with small magnitude of $P$ ($P/\left \langle Q_w \right \rangle ^{0.5} < $ 0.1) for the relatively "incompressible" region of the flow. These points comprise approximately $60\%$ of the flow.
	
	The procedure used to obtain the conditional mean vectors (CMVs) in this study is similar to that in \citet{ooi1999study}. Based on the conditional averages introduced in equation \ref{eqn:condition}, $X(Q, R)$ represents a statistical quantity that is conditioned on $Q$ and $R$. The statistical convergence concerning the bin sizes and the number of samples in each bin has been discussed in section \ref{subsec:converge}.

	 The normalized conditional mean vectors ($DQ/Dt/\left \langle Q_w \right \rangle ^{3/2}$, $DR/Dt/\left \langle Q_w \right \rangle ^{2}$) for different flows are shown in figure \ref{fig:vectorQR}. The vectors obtained from isotropic turbulence data are shown in figure \ref{fig:vectorQR} (a) for reference. It can be seen that the CMVs exhibit a circulating behavior in the $(Q, R)$ plot around the origin in the clockwise direction, indicating that the flow evolves from SFS to UFC, UN/S/S, SN/S/S then back to SFS on average. This circulating behavior represents the Lagrangian dynamics in fully developed turbulence that maintains the tear-drop shape of the ($Q, R$) distribution. This has been observed in many incompressible/compressible canonical turbulent flows \citep{ooi1999study,chu2013topological}. The CMVs for single-fluid and multi-fluid post-shock turbulence are shown in figure \ref{fig:vectorQR} (b) and (c). Evidently, the joint PDF of ($Q, R$) becomes more symmetric due to shock compression. From the Lagrangian point of view, the circulating behavior as seen in figure \ref{fig:vectorQR} (a) for isotropic turbulence is weakened. The particles in $Q_2$ tend to have an increasing $Q$ and decreasing $R$, resulting in an overall trend of getting away from the original point, instead of circulating and then moving toward $Q_1$. This trend in the second quadrant represents an increase of enstrophy.  The particles in $Q_1$ have similar dynamics as in isotropic turbulence and tend to move downward in the ($Q, R$) plane toward the zero discriminant curve. The particles in $Q_3$ are more likely to move straight up towards $Q_2$, while those in $Q_4$ are likely to move away from the original point following the direction of the zero discriminant line and then circulate back to $Q_3$. The overall behavior formed by these particles demonstrates the return-to-isotropy process, with an enlarging head in the second quadrant and elongating tail in the fourth quadrant, anticipating the formation of the classic tear-drop shape. 
	
	\begin{figure}
		\centering
		\includegraphics[width=5in]{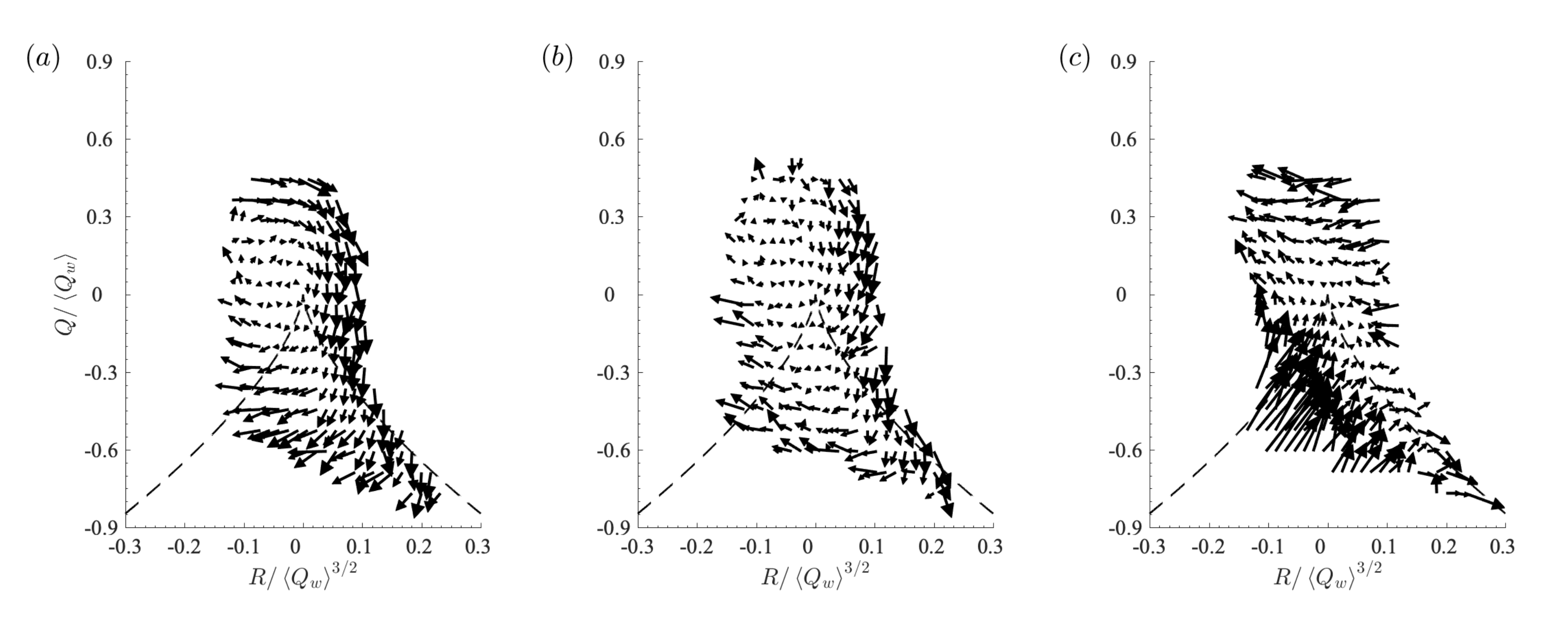}
		\caption{Conditional mean vectors in the ($Q, R$) invariants plane for (\textit{a}) light fluid, (\textit{b}) medium density fluid and (\textit{c}) heavy fluid at streamwise location of $k_0x \approx 0.5$.}
		\label{fig:vectorQRmhl}
	\end{figure}

	The density effects can be further examined by conditioning the ($DQ/Dt, DR/Dt$) vector field on the local density. Figure \ref{fig:vectorQRmhl} (a) shows the CMVs for the light fluid regions. The light fluid particles retain the circulating motion, except that the particles in $Q_3$ and $Q_4$ are likely to go straight left instead of following the zero discriminant line. In general, the flow dynamics in the light fluid regions are less affected by the shock wave. For the medium density fluid regions (figure \ref{fig:vectorQRmhl} b), the circulating motion disappears. On the right side of the ($Q, R$) plane ($R>0$), which is the strong dissipation-production region based on equation \ref{eqn:qrsw}, the fluid particles tend to move downward, resulting in lower $Q$ values.  On the left side of the 
	($Q, R$) plane ($R<0$), which is the enstrophy-production dominated region, the fluid particles tend to move to the left, indicating an increased enstrophy-production. The overall downward-moving behavior of the medium density fluid particles is indicative of decreasing vorticity. This is possibly due to the fact that vorticity is preferentially amplified in the medium density region across the shock wave. After passing the shock wave, the vorticity will decrease as the correlation between density and vorticity vanishes. Figure \ref{fig:vectorQRmhl} (c) shows the CMVs for the heavy fluid regions. Interestingly, the heavy fluid particles exhibit counterclockwise motion. The heavy particles start from $Q_3$ and move to $Q_4$, $Q_1$, and finally to $Q_2$. This implies that they become vorticity dominated due to the fast depletion of strain. Evidently, density plays an important role in the development of the flow topology in the post-shock region, so special attention should be made to the modeling of variable density STI. 
	
 \begin{figure}
		\centering
		\includegraphics[width=4in]{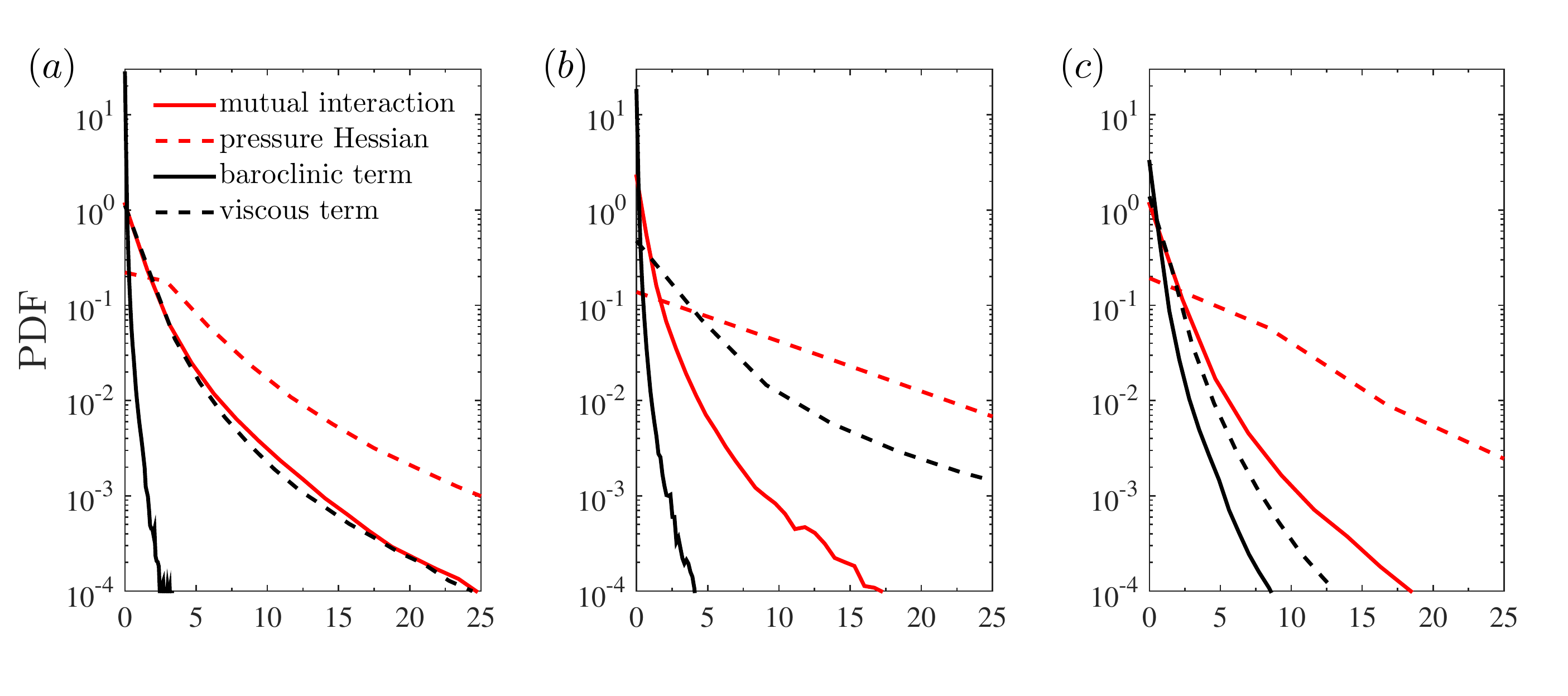}
		\caption{PDFs of the normalized magnitude of the different contributions from Lagrangian dynamics for  (\textit{a}) isotropic turbulence, (\textit{b}) single-fluid post-shock turbulence, and (\textit{c}) multi-fluid post-shock turbulence.}
		\label{fig:cmt_pdf}
	\end{figure}
		
To better understand the underlying mechanisms that cause the behavior highlighted above, the dynamic equations (\ref{eqn:lagdynamic}) governing the vector ($DQ/Dt, DR/Dt$) are examined. In figure \ref{fig:cmt_pdf}, PDFs of the normalized magnitude of the different contributions from Lagrangian equations are shown to study the relative importance of different dynamics. The normalization used here for the vectors is the same as that used in figure \ref{fig:vectorQR}. Figure \ref{fig:cmt_pdf} (a) shows that for isotropic turbulence, the pressure Hessian term has the largest magnitude and the baroclinic contribution is the smallest. Mutual interaction and viscous terms have almost the same magnitude and distribution. After interacting with the shock wave, the magnitude of the baroclinic term is amplified for both single- and multi-fluid turbulence, but still remains the smallest comparing to the other contributions. The mutual interaction term becomes less important due to its reduced magnitude for both cases. The viscous term, however, exhibits different behavior between single- and multi-fluid cases: it is amplified in the single-fluid case and reduced in the multi-fluid case. The pressure Hessian term is also amplified and remains the largest among all the terms. The percentage of contributions, using the means indicate that the percentage of pressure Hessian contribution increases from 61.3\% to 74.9\% (single-fluid) and 73.9\% (multi-fluid) across the shock wave.
		
 \begin{figure}
		\centering
		\includegraphics[width=5in]{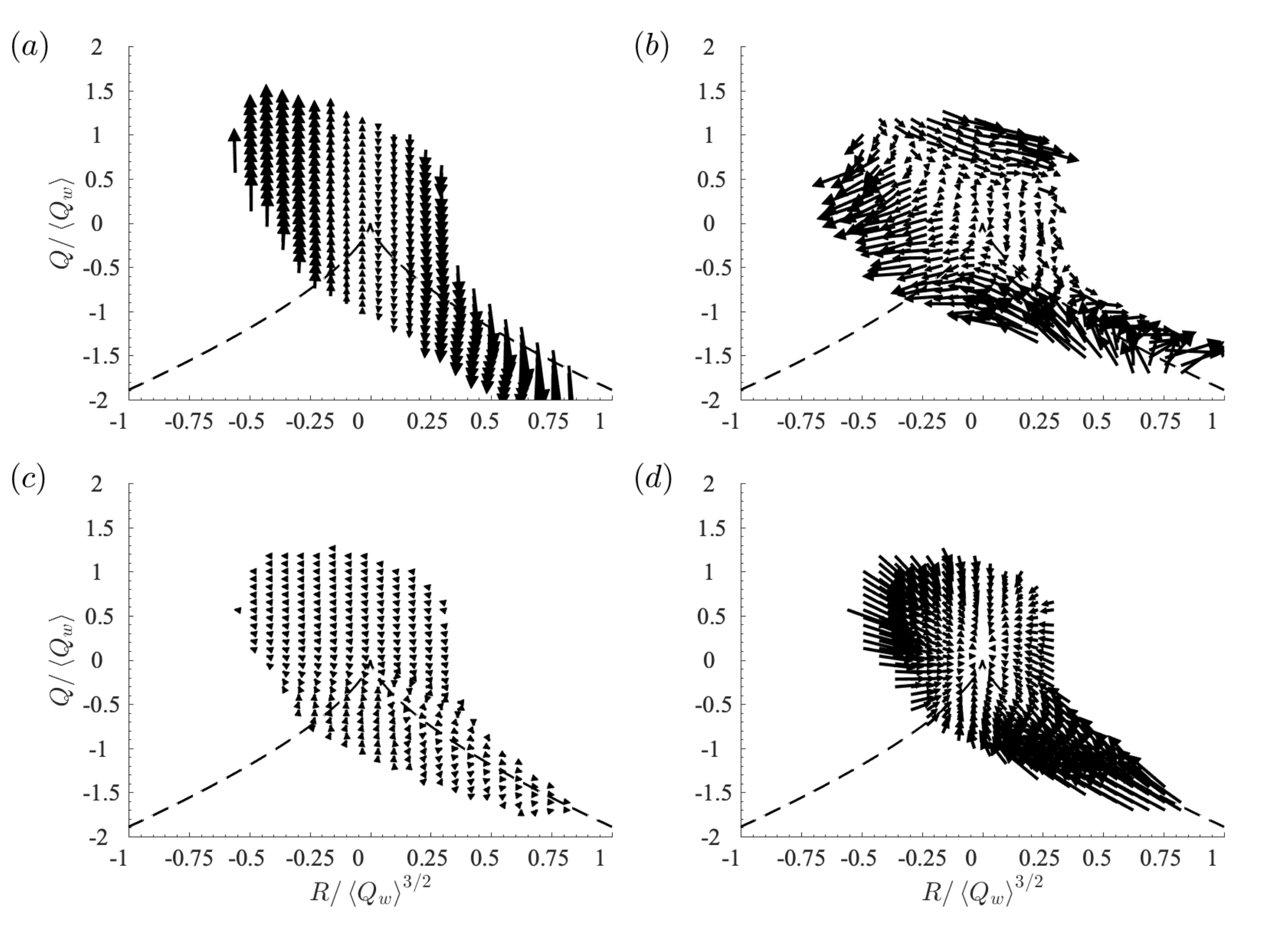}
		\caption{Contributions to the transport equations of the VGT invariants by different terms for isotropic turbulence. (\textit{a}) Mutual interaction among invariants, (\textit{b}) pressure Hessian term, (\textit{c}) baroclinic term, and (\textit{d}) viscous term.}
		\label{fig:vectorspre}
\end{figure}
	
 The Lagrangian dynamics of the flow can be understood better by considering the conditional mean vectors of different terms in the $(Q, R)$ plane. As a reference, these terms are shown in figure \ref{fig:vectorspre} for isotropic turbulence. The variable $Q$ tends to be amplified in the enstrophy-production dominated region due to the effects of vortex stretching mechanism and is decreased in the dissipation-production dominated region due to self-amplification of the strain rate tensor. On the other hand, the mutual effects on $R$ are small because the first invariant $P$ is usually small and the positive and negative values are likely to cancel each other. The contributions from the pressure Hessian (figure \ref{fig:vectorspre} b) tend to move the particles away from an asymptotic line, ending up amplifying the magnitude of $R$. This result agrees well with that observed in turbulent boundary layers \citep{chu2013topological}. For the current simulation, the asymptotic line starts from $Q_2$ and ends in $Q_4$ with a slope of around -2.5. The baroclinic contributions are very small in the post-shock turbulence as shown in figure \ref{fig:vectorspre} (c). The viscous effects as shown in figure \ref{fig:vectorspre} (d) and as expected are reducing the magnitudes of $Q$ and $R$ and pushing the particles towards the origin. This has been observed in various types of turbulence \citep{ooi1999study,chu2013topological}. The combined effects from the four above mechanisms determine the circulating behavior of the conditional mean of ($DQ/Dt, DR/Dt$) vectors.

	\begin{figure}
		\centering
		\includegraphics[width=5in]{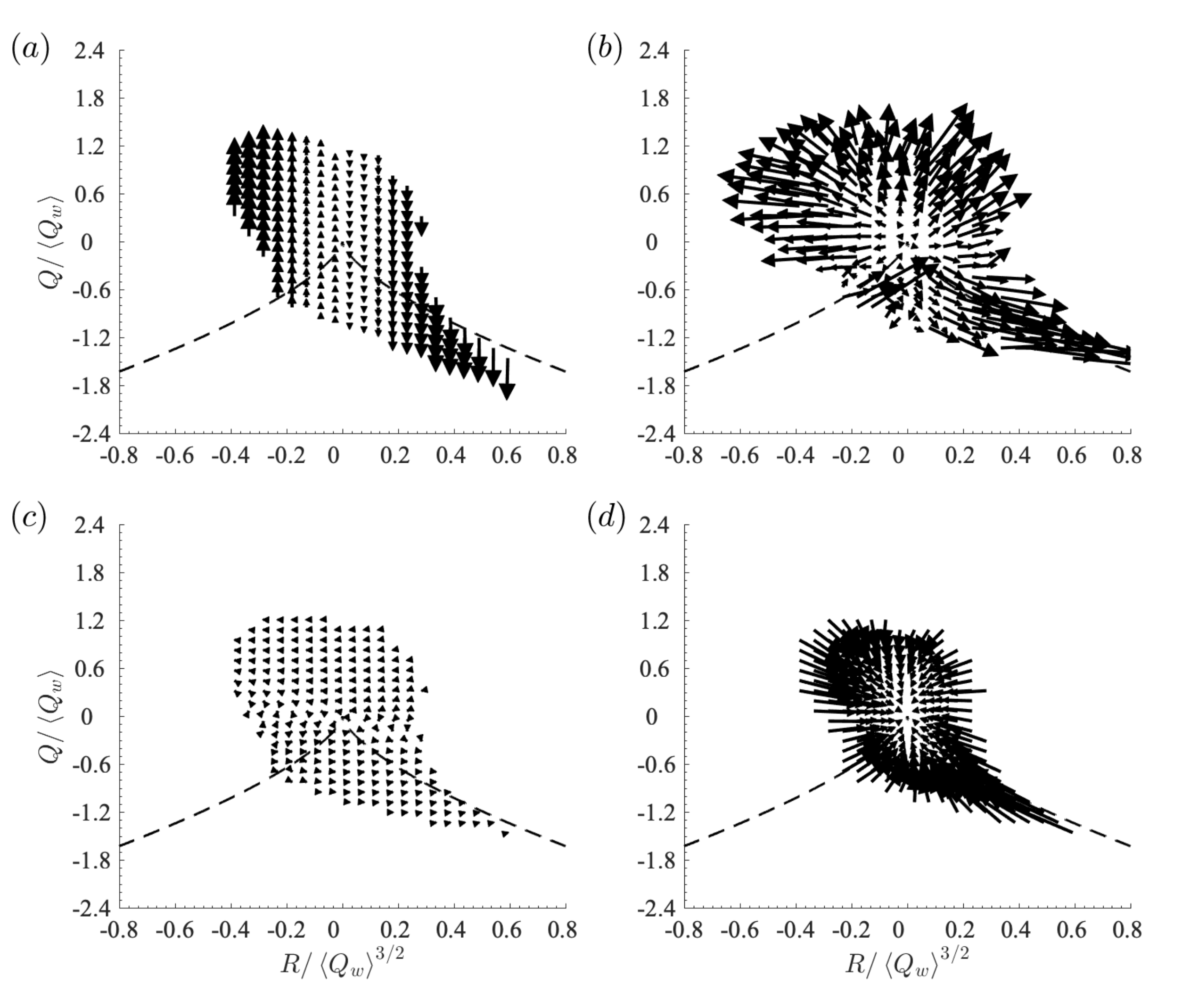}
		\caption{Contributions to the transport equations of the VGT invariants by different terms for single-fluid post-shock turbulence. (\textit{a}) Mutual interaction among invariants, (\textit{b}) pressure Hessian term, (\textit{c}) baroclinic term, and (\textit{d}) viscous term.}
		\label{fig:vectorspost}
	\end{figure}
	
	After interaction with the shock wave, the conditional mean vectors in the ($Q, R$) plane are different from those in the pre-shock isotropic turbulence. Figure \ref{fig:vectorspost} shows the results for the single-fluid case. By comparing it with figure \ref{fig:vectorspre}, we note that even though the conditional means of ($DQ/Dt, DR/Dt$) vectors are different, the contributions from mutual interaction (figure \ref{fig:vectorspost} a), baroclinic term (figure \ref{fig:vectorspost} c) and viscous term (figure \ref{fig:vectorspost} d) are very similar. The only term that is qualitatively different in the post-shock turbulence and isotropic turbulence is the pressure Hessian term (figure \ref{fig:vectorspost} b). The importance of the pressure Hessian term is also reflected on the dynamical contributions in the $(Q, R)$ plane. In the post-shock single-fluid turbulence, the asymptotic line that separates the vectors into two regions with different behaviors disappears. Instead, the pressure Hessian term tends to move the particles away from the origin in $Q_1$ and $Q_2$, thus increasing Q and $|R|$ values of the particles. In $Q_3$ and $Q_4$, the vectors are parallel to the left zero discriminant line, making the particles move from $Q_3$ to $Q_4$, and then to $Q_1$.
	
\begin{figure}
		\centering
		\includegraphics[width=5in]{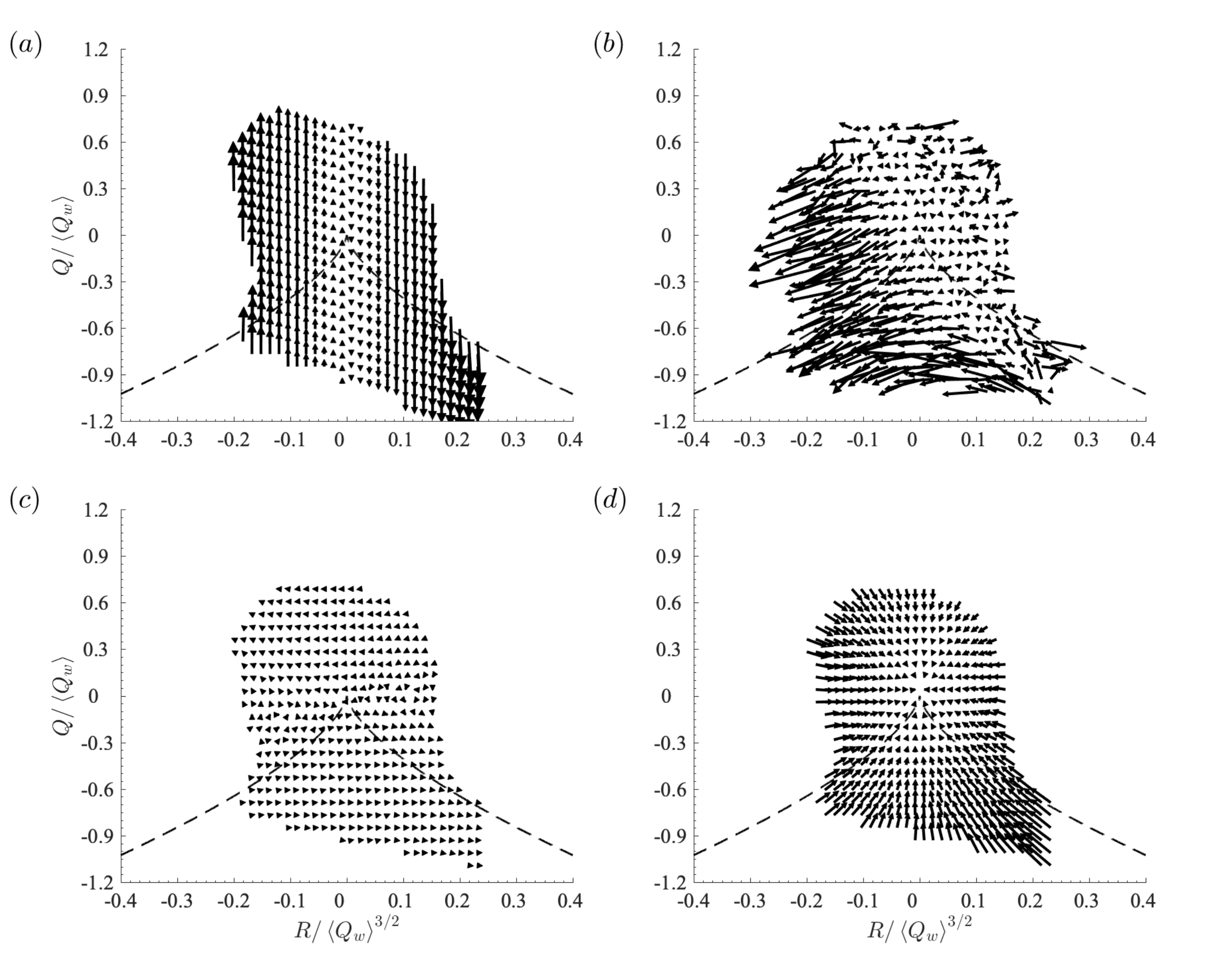}
		\caption{Contributions to the dynamics of the VGT invariants by different terms for multi-fluid post-shock turbulence. (\textit{a}) Mutual interaction among invariants, (\textit{b}) pressure Hessian term, (\textit{c}) baroclinic term, and (\textit{d}) viscous term.}
		\label{fig:vectormpost}
	\end{figure}

\begin{figure}
		\centering
		\includegraphics[width=5in]{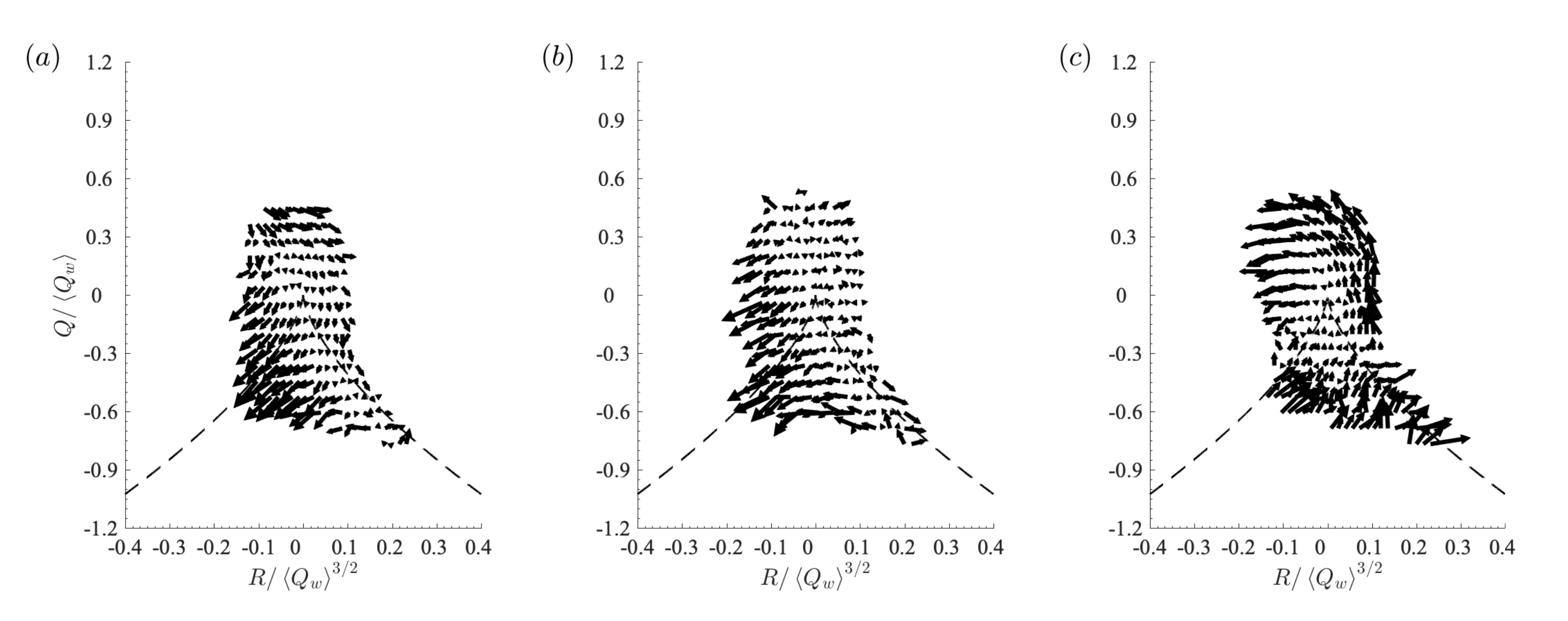}
		\caption{Contributions from pressure Hessian to the dynamics of the VGT invariants in (\textit{a}) light fluid region, (\textit{a}) medium density fluid region and (\textit{c}) heavy fluid region.}
		\label{fig:vectort2mhl}
	\end{figure}

	For multi-fluid post-shock turbulence, the pressure Hessian term is also the only term that is qualitatively different than that in isotropic turbulence (figure \ref{fig:vectormpost}). Despite the increased density and pressure gradient in the multi-fluid case, the baroclinic term is still considerably smaller than all the other terms. In $Q_2$ and $Q_3$, an asymptotic line similar to that in isotropic turbulence seems to exist, which "repels" the vectors away from it, causing an increase in $|R|$ values. In $Q_1$ and $Q_4$, the magnitude of pressure hessian term becomes much smaller. The further conditioned pressure Hessian term based on the local densities in figure \ref{fig:vectort2mhl} indicates that fluid particles with different densities have very different behaviors with respect to pressure Hessian dynamics. Specifically, the pressure Hessian generally moves the heavy particles toward the regions with larger $Q$ values. In $Q_3$ and $Q_4$, it also moves the heavy fluid particles towards the $R>0$ plane. For the light fluid particles, the pressure Hessian term tends to make them move towards regions with larger $|R|$ values in the first and second quadrant. In $Q_3$ and $Q_4$, the fluid particles move from $Q_4$ to $Q_3$. Last but not the least, the fluid particles with medium density seem to exhibit similar behavior to light fluid particles, except in $Q_1$, where the pressure Hessian contribution is moving the fluid particles towards the regions with large Q values. Examining figure \ref{fig:vectorQRmhl} and figure \ref{fig:vectort2mhl} together, we observe that the differences in particle dynamics in the ($Q, R$) plane in regions with different densities are mainly due to differences in the pressure Hessian contributions. 
	
	\section{Conclusions}
	\label{sec:conclusion}
	
	Accurate shock-capturing turbulence-resolving simulations are used together with Eulerian and Lagrangian particle tracking post-processing methods to investigate the interaction of an isotropic turbulence with a normal shock wave for both single-fluid and a binary mixture of different density fluids (species). The main objective is to develop a better understanding of the variable density effects on the post-shock turbulence structure and its evolution away from the shock. Grid convergence tests are conducted to establish the numerical accuracy of the simulated data. The results show that the turbulence statistics are grid-converged, indicating good accuracy of the current computational method. Statistical convergence is also conducted for Lagrangian data.
	
	The analysis is restricted here to an Atwood number of 0.28, based on the molar masses of the two fluids. At this Atwood number value, the variable density effects introduce important changes in the turbulence structure, while the shock remains in the wrinkled regime for the shock Mach number considered. Similarly, the turbulent Mach number is also small enough that the multi-fluid case does not transition to the broken shock regime and the post-shock compressibility effects are weak. On the other hand, the Reynolds number is large enough so that the viscous effects stay small through the interaction with the shock and the corresponding single-fluid simulation satisfies the LIA limit. As the flow transitions to the broken shock regime due to larger turbulent Mach number and/or Atwood number, additional effects appear. These are left for future studies.
	
	The density effects on the post-shock turbulence structure are first studied using Eulerian data. The different roles that the variable density field could play through the shock interaction are identified for some of important statistics. The non-Gaussianity of the velocity gradient tensor (VGT) is studied by examining the PDFs of velocity gradient components. The preferential amplification of rotation and strain rate tensors is found to be affected by the density variations, leading to a weaker correlation between the two tensors in the multi-fluid case. This is shown to be caused by different roles that density plays on the modification of rotation and strain rate tensors across the shock wave. The skewness and flatness of VGT components before and after the shock wave are then examined to study the evolution of VGT. It is shown that density effects are weak across the shock, but are stronger in the post-shock development. The density variations are also shown to cause the preferential alignment between eigenvectors of the strain rate tensor and density gradient vector, which then modifies the skewness of the velocity gradient and density gradient PDFs. 
	
	 The density effects on the flow topology are then examined by first comparing the joint PDF of the second and third invariants of the deviatoric part of the velocity gradient tensor. The pre-shock joint PDF has the classic tear-drop shape. However, after the shock wave, a tendency towards symmetrization of the joint PDF, as in single-fluid STI, is observed for the multi-fluid case, with more data points falling into the first and third quadrants. After conditioning the joint PDFs based on fluid density, large differences among heavy, medium, and light fluid regions are observed. In the heavy fluid regions, the joint PDF becomes almost completely symmetrical with an increasing portion of data fall in the third quadrant. In contrast, the majority of the light fluid data points have a similar distribution to that of isotropic turbulence. A connection between low vortex stretching and the joint $Q^\ast$,$R^\ast$ statistics is established for the post-shock turbulence, by considering the contribution to vortex stretching rate from each quadrant.
	
	Furthermore, Lagrangian fluid particles are used to track the development of the turbulence and VGT after the interaction with the shock. The Lagrangian dynamics of the VGT are also examined by using the conditional mean rate of change of the invariants of VGT. For the parameter range considered, the results show that particles in $Q_3$ have the least residence time, while those in $Q_2$ have the longest residence times. The residence times are smaller than those in isotropic turbulence, especially in the multi-fluid case. It is also shown that particles starting in quadrants $Q_1$ and $Q_3$ play an important role in recovering of the vortex stretching term. After interacting with the shock wave, the "clockwise circulating" behavior (as observed in the isotropic turbulence) disappears in both single- and multi-fluid cases. Our analysis highlights the mechanisms through which post-shock turbulence recovers the classical tear-drop shape, with an enlarging head in the second quadrant and elongating tail in the fourth quadrant. The contributions from different terms in the dynamic equations of VGT invariants, compared with isotropic turbulence, show that the pressure Hessian term is critical to the topological evolution of turbulence. The relative magnitude of the pressure Hessian term is increased and its dynamical contributions in $(Q, R)$ plane are modified across the shock wave. The pressure Hessian term is also shown to be strongly dependent on the local density in the multi-fluid case, resulting in completely different dynamics in regions with different densities. In this work, the out-of-plane $(Q, R)$ compressibility effects are not considered due to the relatively low $M_t$ and $A_t$. The compressibility effects and their coupling with the variable density effects will be investigated in more detail in future studies.

	\section*{Acknowledgements}
	This work was performed under the auspices of DOE. YT and FAJ were supported by Los Alamos National Laboratory, under Grant No. 319838. Los Alamos National Laboratory, an affirmative action/equal opportunity employer, is managed by Triad National Security, LLC, for the National Nuclear Security Administration of the U.S. Department of Energy under contract 89233218CNA000001. Computational resources were provided by the High Performance Computing Center at Michigan State University and Texas Advanced Computing Center (TACC) at The University of Texas at Austin. 
	
	\bibliographystyle{jfm}
	% Note the spaces between the initials
	\bibliography{jfm}

\end{document}